\documentclass{article}
\usepackage{fullpage}
\usepackage{authblk}
\usepackage{amsmath}
\usepackage{amssymb}
\usepackage{graphicx}
\usepackage[colorlinks=true]{hyperref}
\usepackage[font=small,labelfont=bf]{caption}

\begin{document}

\title{Improved Volume-of-Solid formulations for micro-continuum simulation of mineral dissolution at the pore-scale}

\author[1]{Julien Maes}
\author[2]{Cyprien Soulaine}
\author[1]{Hannah P. Menke}
\affil[1]{Institute of GeoEnergy Engineering, Heriot-Watt University, Edinburgh, U.K.}
\affil[2]{CNRS, University of Orl\'eans, Orl\'eans, France}

\maketitle

\begin{abstract}
We present two novel Volume-of-Solid (VoS) formulations for micro-continuum simulation of mineral dissolution at the pore-scale. The traditional VoS formulation (VoS-$\psi$) uses a diffuse interface localization function $\psi$ to ensure stability and limit diffusion of the reactive surface. The main limitation of this formulation is that accuracy is strongly dependent on the choice of the localization function. Our first novel improved formulation (iVoS) uses the divergence of a reactive flux to localize the reaction at the fluid-solid interface, so no localization function is required. Our second novel formulation (VoS-$\psi$') uses a localization function with a parameter that is fitted to ensure that the reactive surface area is conserved globally. Both novel methods are validated by comparison with experiments, numerical simulations using an interface tracking method based on the Arbitrary Eulerian Lagrangian (ALE) framework, and numerical simulations using the VoS-$\psi$. All numerical methods are implemented in GeoChemFoam, our reactive transport toolbox and three benchmark test cases in both synthetic and real pore geometries are considered: (1) dissolution of a calcite post by acid injection in a microchannel and experimental comparison, (2) dissolution in a 2D polydisperse disc micromodel at different dissolution regimes and (3) dissolution in a Ketton carbonate rock sample and comparison to \textit{in-situ} micro-CT experiments. We find that the iVoS results match accurately experimental results and simulation results obtained with the ALE method, while the VoS-$\psi$ method leads to inaccuracies that are mostly corrected by the VoS-$\psi$' formulation. In addition, the VoS methods are significantly faster than the ALE method, with a speed-up factor of between 2 and 12.   

\end{abstract}

\section{Introduction}

\label{Sect:Intro}
Prediction of solid mineral dissolution during reactive flow in porous media is vital for a wide range of subsurface applications, including CO$_2$ sequestration \cite{2015-Black}, geothermal systems \cite{2015-Pandey} and enhanced oil recovery
\cite{2017-Shafiq}. CO$_2$ storage in underground reservoirs has the potential to significantly mitigate the environmental impact of many industrial processes. However, mineral dissolution is a potential barrier to the long-term
storage of CO$_2$ in the subsurface, as CO$_2$ reacts with water to make carbonic acid that can dissolve solid minerals and threaten the structural integrity of a reservoir \cite{2011-Nordbotten}. In addition, most subsurface applications involve the injection of fluids with chemical properties that are incompatible with existing reservoir fluids and can lead to mineral precipitation or scaling in the pore structure. Scaling is especially prevalent near well-bores, and can significantly reduce the
permeability, and thus productivity of a porous formation. Acid injection is then often used to improve the flow in clogged wells \cite{1979-Williams}. Thus, accurate and efficient modelling
of mineral dissolution in porous media is crucial to improve and optimise these engineering processes.

Modelling reactive transport at the field-scale relies on the assumption that a representative elementary volume can be defined such that flow, transport and reaction can be described in terms of bulk properties like porosity, permeability and macro-scale reactive constant, in what is usually referred as the Darcy scale \cite{1994-Quintard,1988-Shapiro,1988-Lichtner,2015-Steefel}. Mineral dissolution modifies the pore structure and results in a change in these Darcy-scale properties. Pore-scale numerical experiments can be used to predict the change in these Darcy-scale properties during dissolution. At the pore-scale, these reactions are applied directly on the solid surface while resolving flow and transport in a representative elementary volume of pore space directly. The effects of these dissolution-induced structural changes on the flow and transport properties of the bulk medium can then be estimated for use in Darcy-scale simulations.

The last decade has seen an explosion in the study of flow and transport behaviour at the pore-scale \cite{2009-Noirel,2017-Menke,2013-Raoof,2013-Varloteaux,2017-Molins,2016-Starchenko,2017-Soulaine,2020-Yang}. Recent advances in X-ray imaging techniques have enabled direct observation and quantification of dissolution-induced changes in pore structures \cite{2009-Noirel,2013-Hao,2014-Luquot,2015-Menke,2016b-Menke,2017-Menke,2018-Menke}. Numerical modeling has played an important role in these investigation of pore-scale physics, as it provides a mechanistic understanding of the relevant coupled processes. Furthermore, simulation results resolve variables that are not easily available from experiments such as concentration gradients within the pore space.

Numerical modelling of mineral dissolution at the pore-scale can be performed using Pore-Network Modelling \cite{2013-Raoof,2013-Nogues,2013-Varloteaux}. However, the evolution
of the pore-space can only be predicted using the finite range of geometrical parameters of the network. Alternatively, computational microfluidics \cite{2021-Soulaine} has been applied using a range of numerical methods \cite{2020-Molins}. Interface tracking models explicitly deform and move the solid surface, either using solid balance with a threshold on a lattice \cite{2002-Kang,2009-Szymczak,2011-Prasianakis}, a conforming mesh based on the Arbitrary-Lagrangian-Eulerian (ALE) framework \cite{2016-Starchenko,2018-Starchenko} or smoothed particle hydrodynamics \cite{2007-Tartakovsky}. Alternatively, the interface can be captured using a level-set function \cite{2010-Li,2017-Molins}. For all these methods, the boundary conditions on the solid surface can be be applied directly, or using an immersed boundary condition. However, they require additional treatment for interface displacement, topological changes or remeshing, which usually lead to an increase in their computational cost \cite{2021-Yang}.

The micro-continuum approach \cite{2016a-Soulaine,2017-Soulaine,2019-Soulaine,2016-Chatelin,2021-Yang} based on the Volume-of-Solid (VoS) method offers an attractive substitute for interface tracking models. Within this approach, the fluid-solid interface is captured using an indicator function equal to the volume fraction of void space in each cell, and flow and transport are solved using the Darcy-Brinkman-Stokes (DBS) equation. The VoS method is computationally efficient as it does not require remeshing or any special treatment for topological changes. 

In the standard VoS approach, the surface area of the fluid-solid interface in a control volume is computed through the gradient of a volume fraction. In practice, a diffuse interface may emerge that spreads across a large number of layers in the computational grid. To enforce the localization of the reactive boundary condition at the fluid/solid interface, a diffuse interface localization function $\psi$ is generally introduced \cite{2017-Soulaine} and this formulation is labelled VoS-$\psi$. The main advantage of the VoS-$\psi$ method is that standard Reactive Transport Modelling dedicated to Darcy-scale can be easily applied to simulate geochemical processes at the pore-scale by simply changing the way the fluid-rock interfacial area within control volume is estimated \cite{2021b-Soulaine}. While the surface area for Darcy-scale simulations is an input parameter that is either constant or depends on complex function of porosity and flow rates \cite{2009-Noirel, 2017-Wen}, the surface area for pore-scale simulations using VoS is directly calculated from the mapping of the solid volume fraction. The main limitation of the VoS$-\psi$ is that the accuracy of the model depends strongly on the choice of the localization function \cite{2012-Luo}, and the optimal choice depends on a large number of parameters, such as the geometry, the flow rate, the reactive constant, the computational mesh and the discretization method used for the computation of gradients.

In this paper we propose two novel VoS formulations. The first formulation (iVoS) removes the need for a localization function by computing the reaction rate using the divergence of flux. The second formulation (VoS-$\psi$') uses a localization function with a parameter that is fitted to ensure that the reactive surface area is conserved globally. The numerical models are presented in Section \ref{Sect:model}. The iVoS and VoS-$\psi'$ methods are then compared with the VoS-$\psi$ method and with an interface tracking method based on the ALE framework on three benchmark test cases in Section \ref{Sect:benchmark}. In each case, we show that our new VoS methods match accurately experimental results and/or simulation results using the ALE method while being significantly faster.  

\section{Mathematical models}

In this section, the governing equations and the micro-continuum approach are first presented. Then, we describe the iVoS, VoS-$\psi$ and VoS-$\psi$' models, which differ only in the way the reaction rate is computed. Further, we show how dimensionless analysis lead to the quasi-static assumption that reduces the computational time.

\label{Sect:model}

\subsection{Governing equations}

The governing equations consider flow, transport and reaction at the fluid-solid interface. The domain $\Omega$ is partitioned into fluid $\Omega_f$ and solid $\Omega_s$. Under isothermal conditions and in the absence of gravitational effects, fluid motion in $\Omega_f$ is governed by the incompressible Navier-Stokes equations
\begin{equation}\label{Eq:cont}\nabla\cdot\mathbf{u} = 0,
\end{equation}
\begin{equation}
\frac{\partial \mathbf{u}}{\partial t}+ \nabla\cdot\left(\mathbf{u}\otimes\mathbf{u}\right)=-\nabla p +\nu\nabla^2\mathbf{u},\label{Equ:momentum}
\end{equation}
with the continuity condition at the fluid-solid interface $\Gamma$,
\begin{equation}\label{Equ:bcu}
\rho\left(\mathbf{u}-\mathbf{w}_s\right)\cdot \mathbf{n}_{s}=-\rho_s\mathbf{w}_s\cdot\mathbf{n}_s \hspace{0.5cm} \text{at $\Gamma$},
\end{equation}
where $\mathbf{u}$ (m/s) is the velocity, $p$ (m$^2$/s$^2$) is the kinematic pressure, $\nu$ (m$^2$/s) is the kinematic viscosity, $\rho$ (kg/m$^3$) is the fluid density, $\rho_s$ (kg/m$^3$) is the solid density, $\mathbf{n_s}$ is the normal vector to the fluid-solid interface pointing toward the solid phase, and $\mathbf{w_s}$ (m/s) is the velocity of the fluid-solid interface, which is controlled  by the surface reaction rate $R$ (kmol/m$^2$/s) such that
\begin{equation}\label{Eq:Ws}
\mathbf{w}_s=\frac{M_{ws}}{\rho_s}R\mathbf{n}_s,
\end{equation}
where $M_{ws}$ is the molecular weight of the solid. The concentration $c$ (kmol/m$^3$) of a species in the system satisfies an advection-diffusion equation
\begin{equation}\label{Eq:concentration}
\frac{\partial c}{\partial t}+ \nabla \cdot \left( c\mathbf{u} \right) = \nabla\cdot\left(D\nabla c\right),
\end{equation}
where  $D$ (m$^2$/s) is the diffusion coefficient. The chemical reaction occurs at the fluid-solid interface $\Gamma$, such that
\begin{equation}\label{Equ:bcc1}
\left(c\left(\mathbf{u}-\mathbf{w}_s\right)-D\nabla c \right)\cdot \mathbf{n}_{s}=\zeta R \hspace{0.5cm} \text{at $\Gamma$},
\end{equation}
where $\zeta$ is the stoichiometric coefficient of the species in the reaction. In this work, we assume that the surface reaction rate depends only on the concentration of one reactant species, following
\begin{equation}
 R=k_cc,
\end{equation}
where $k_c$ (m/s) is the reaction constant.

\subsection{Micro-continuum approach with Volume-Of-Solid}
In the micro-continuum approach, the entire domain $\Omega$ is considered, i.e fluid $\Omega_f$ and solid $\Omega_s$, and the fluid-solid interface is tracked in terms of $V_f$ and $V_s$, the volume of fluid and solid phase in each control volume $V$, and their volume fraction $\varepsilon=V_f/V$ and $\varepsilon_s=1-\varepsilon$. The flow, transport and chemical reaction are solved in term of the volume-averaged velocity
\begin{equation}
 \overline{\mathbf{u}}=\frac{1}{V}\int_{V_f}\mathbf{u}dV,
\end{equation}
and the phase-averaged pressure and reactant concentration
\begin{align}
 &\overline{p}=\frac{1}{V_f}\int_{V_f}pdV,\\
 &\overline{c}=\frac{1}{V_f}\int_{V_f}cdV.
\end{align}
The averaging process results in an extension of the Darcy-Brinkman-Stokes equation \cite{2017-Soulaine}
\begin{equation}\label{Eq:momentumDBS}
 \frac{1}{\varepsilon}\left(\frac{\partial \mathbf{u}}{\partial t}+\nabla\cdot\left(\frac{\mathbf{u}\otimes\mathbf{u}}{\varepsilon}\right)\right)=-\nabla \overline{p} +\frac{\nu}{\varepsilon}\nabla^2\overline{\mathbf{u}}-\nu k^{-1}\overline{\mathbf{u}},
\end{equation}
where $k$ (m$^2$) is the permeability of the cell. $\nu k^{-1}\overline{\mathbf{u}}$ represents the momentum exchange between the fluid and the solid phase, i.e. the Darcy resistance. This term is dominant in the solid phase and vanishes in the fluid phase. To model this, the local permeability field $k$ is assumed to be a function of the local porosity $\varepsilon$, following a Kozeny-Carman relationship
\begin{equation}\label{Eq:perm}
 k=k_0\frac{\varepsilon^3}{\left(1-\varepsilon\right)^2},
\end{equation}
where $k_0$ (m$^{2}$) is the Kozeny-Carman constant. For the acid transport, the mass-balance equation averaged over the control volume gives
\begin{equation}\label{Eq:concentrationDBS}
 \frac{\partial \varepsilon_f\overline{c}}{\partial t}+\nabla\cdot\left(\overline{c}\overline{\mathbf{u}}\right)-\nabla\cdot\left(\varepsilon D^*\nabla\overline{c}\right)+\overline{R}_f=0,
\end{equation}
where $\varepsilon D^*$ (m$^2$/s) is the effective diffusion coefficient and $\overline{R}$ (kmol/m$^3$/s) is the volume-averaged surface reaction rate. The effective diffusion coefficient takes into account the reduction of the total diffusion due to the presence of solid phase. In this paper, we take $D^*=D$. The volume-averaged surface reaction rate is defined as
\begin{equation}
\overline{R}=\frac{1}{V}\int_{A}k_ccdS,
\end{equation}
where $A=V\cap\Gamma$ is the reactive surface area in the control volume. The specific surface area $a_s$ (m$^{-1}$) in a control volume is defined as
\begin{equation}
 a_s=\frac{1}{V}\int_{A}dS.
\end{equation}
Finally, the mass balance equation for the solid phase writes
\begin{equation}\label{Eq:mbs}
 \frac{\partial \varepsilon}{\partial t}=\overline{R}\frac{M_{ws}}{\rho_s}.
\end{equation}
and the volume averaged velocity satisfies
\begin{equation}\label{Eq:contDBS}
 \nabla\cdot\overline{\mathbf{u}} = \overline{R}M_{ws}\left(\frac{1}{\rho_s}-\frac{1}{\rho}\right).
\end{equation}

\subsection{Improved Volume-of-Solid}
The method presented here is analogue to the calculation of the mass transfer across a multiphase interface presented in \cite{2020-Maes}, for which the mass transfer is calculated as the scalar product between a diffusive flux and the gradient of the phase indicator function.
To calculate the volume-averaged surface reaction rate, the improved Volume-of-Solid (iVoS) therefore introduces the reactive flux, $\mathbf{\Phi}_R$ (kmol/m$^2$/s), defined as
\begin{equation}
 \mathbf{\Phi}_R=k_cc\mathbf{n}_s,
\end{equation}
and the volume-averaged surface reaction rate can be rewritten as
\begin{equation}
\overline{R}=\frac{1}{V}\int_{A}\mathbf{\Phi}_R\cdot\mathbf{n}_sdS.
\end{equation}
Assuming that the concentration of the reactant on the reactive surface can be approximated by its volume-averaged on the control volume, and that the normal vector to the interface can be approximated by
\begin{equation}
\overline{\mathbf{n}}_s=-\frac{\nabla\varepsilon}{\|\nabla\varepsilon\|},
\end{equation}
the reactive flux can be approximated by
\begin{equation}
 \overline{\mathbf{\Phi}}_R=k_c\overline{c}\overline{\mathbf{n}}_s.
\end{equation}
Moreover, the average surface normal in a control volume can be calculated as \cite{1994-Quintard}
\begin{equation}
\frac{1}{V}\int_{A}\mathbf{n}_sdS=-\nabla\varepsilon.
\end{equation}
Therefore, the volume-averaged surface reaction rate can be calculated as
\begin{equation}
 \overline{R}=-\overline{\mathbf{\Phi}}_R\cdot\nabla\varepsilon.
\end{equation}
To avoid problems related to the calculation of the gradient of $\varepsilon$ (see section \ref{sec:vospsi}), the divergence theorem is used to recast $\overline{R}$ as
\begin{equation}\label{Eq:RfiVoS}
 \overline{R}=\varepsilon\nabla\cdot\overline{\mathbf{\Phi}}_R-\nabla\cdot\left(\varepsilon\overline{\mathbf{\Phi}}_R\right),
\end{equation}
With this formulation, the reactive rate is the sum of two terms, an overall mass transfer term ($\sum\overline{R}_f=\sum\left(\varepsilon\nabla\cdot\overline{\mathbf{\Phi}}_R\right)$) and a conservative term ($-\sum\left(\nabla\cdot\left(\varepsilon\overline{\mathbf{\Phi}}_R\right)\right)=0$), which balances the local reaction rate between two adjacent control volumes. This means that the reaction can consume reactant in one cell and use it to dissolve solid in a neighbor cell. By using a second-order high resolution difference scheme \cite{1974-vanLeer}, the reaction rate is balanced toward the reactive surface and the diffusion of the solid interface is limited. This is an accurate representation of a reaction at an interface between two cells with $\varepsilon\approx0$ and $\varepsilon=1$, where the reaction rate is calculated using the concentration in the fluid where $\varepsilon=1$ and the reaction dissolves the solid where $\varepsilon\approx0$. This formulation is labelled iVoS.

\subsection{Volume-of-Solid with localization function}
As an alternative to Equ. (\ref{Eq:RfiVoS}), the volume-averaged surface reaction rate can be calculated as
\label{sec:vospsi}
\begin{equation}
 \overline{R}=k_c\overline{c}a_s.
\end{equation}
The specific surface area can be directly calculated as $a_s=~\|\nabla \varepsilon\|$. However, this can lead to a diffuse interface that spreads across a large number of layers in the computational grid. To enforce localization of the dissolution front on the fluid-solid interface, the VoS-$psi$ method introduces a diffuse interface localization function $\psi$ \cite{2017-Soulaine} so that
\begin{equation}\label{Eq:RfVoSpsi}
 \overline{R}=k_c\overline{c}\psi\|\nabla\varepsilon\|.
\end{equation}
The main advantage of VoS-$\psi$ compared to iVoS is that it provides a direct calculation of the reactive surface area in a control volume. Therefore, the reaction rate can be calculated by a dedicated geochemical solver, such as Phreeqc \cite{PARKHURST2015176} or Reaktoro \cite{2017-Walsh}, as it is often done for standard Reactive Transport Modelling dedicated to multi-scale applications \cite{2021b-Soulaine}.

While the VoS-$\psi$ method has proved to be a fast and flexible method to match experimental results \cite{2020-Molins}, this formulation has one main limitation. It is strongly dependent on the choice of the localization function $\psi$. Several functions have been proposed by \cite{2012-Luo}, but their accuracy depends on the case considered and on the discretization scheme used for the gradient.  For example, using a centered difference scheme requires $\psi(1.0)=0$ for stability. However, using a decentered scheme (in the direction of $\varepsilon=0$ to avoid instabilities) will result in a higher and more diffuse reaction rate, due to a higher reactant concentration away from the surface and a larger specific surface area when $\varepsilon=0$. Centered difference schemes are less diffuse, but they result in incomplete dissolution, since a cell with $\varepsilon<1$ but $\varepsilon=1$ for all its neighbors will have a zero reaction rate. Currently, there is no consensus on the ideal combination of localization function and discretization scheme to use for every scenario, as this will depend on the geometry and flow conditions. 

In this work, we use a centered difference scheme for the gradient and $\psi=\lambda\varepsilon(1-\varepsilon)$, which is the most accurate combination of discretization scheme and localization function proposed by \cite{2012-Luo} for the case of dissolution of a calcite post by acid injection \cite{2017-Soulaine}. Typically, $\lambda=4$, but since $\psi\leq1.0$, this will lead to a reduction in interfacial area. For this reason, the VoS-$\psi$ method gives an overall lower dissolution rate that the iVoS method. Instead, $\lambda$ can be calculated as a function of $\varepsilon$ so that
\begin{equation}
\lambda\left(\varepsilon\right)=\frac{\int_{\Omega}\|\nabla\varepsilon\|}{\int_{\Omega}\|\nabla\varepsilon\|\varepsilon\left(1-\varepsilon\right)}.
\end{equation}
In this work, we label VoS-$\psi$ the formulation using $\psi=4\varepsilon\left(1-\varepsilon\right)$ and VoS-$\psi$' the formulation using $\psi=\lambda\left(\varepsilon\right)\varepsilon\left(1-\varepsilon\right)$. Using the VoS-$\psi$' formulation, the total surface area is conserved globally. However, it is not conserved locally. At an interface between two cells with $\varepsilon\approx0$ and $\varepsilon=1$, the interface area will remain close to 0 in both cells even after the correction. 

\subsection{Upscaling to the Darcy scale}

In order to investigate the capabilities of our numerical model to calculate upscaled properties for macro-scale simulations, the flow and reaction in the whole domain are characterised by the the total porosity $\phi$, the Darcy velocity $U_D$ (m/s),
\begin{equation}
 U_D = \frac{Q_i}{A_i},
\end{equation}
and the permeability $K$ (m$^2$),
\begin{equation}
 K=\frac{\nu U_DL_D}{\Delta P}.
\end{equation}
where $Q_i$ (m$^3$/s) is the inlet flow rate, $A_i$ (m$^2$) is the inlet area, 
$L_D$ (m) is the distance between the inlet and outlet, $\Delta P$ (m$^2$/s$^2$) is the kinematic pressure drop. In addition, the chemical reaction is characterised by the total specific surface area $a$ (m$^{-1}$)
\begin{equation}
a=\frac{1}{V_{\Omega}}\int_{\Gamma}dS,
\end{equation}
and the correction factor
$\alpha$, that represents the reduction of the reactive surface area accessible to reactant, and is defined as
\begin{equation}
 \alpha=\frac{1}{a}\frac{\int_{\Gamma}cdS}{\int_{\Omega_f}cdV}.
\end{equation}

Change in effective upscaling parameters $K$, $a$ and $\alpha$ are often modelled as a function of porosity with power law functions \cite{2017-Wen,2019-Seigneur,2009-Noirel}. The parameters of these power-law functions depend strongly on the flow, transport and reaction conditions, which are characterized by the P\'eclet number
\begin{eqnarray}
Pe=\frac{UL}{D},
\end{eqnarray}
which quantifies the relative importance of advective and diffusive transport, and the Damk\"ohler number
\begin{eqnarray}
Da=\frac{k_c}{U},
\end{eqnarray}
which quantifies the relative importance of chemical reaction and advective transport. Here $U$ and $L$ are the reference velocity and length. The product of the Damk\"ohler and P\'eclet numbers is also a relevant quantity called the Kinetic number, defined as
\begin{eqnarray}
Ki=DaPe=\frac{k_cL}{D}.
\end{eqnarray}
In addition, the reactant strength is characterized by
\begin{eqnarray}
 \beta=\frac{c_{i}M_{ws}}{\zeta\rho_s},
\end{eqnarray}
where $c_{i}$ is the concentration of reactant at the inlet.

\section{Benchmark cases}
\label{Sect:benchmark}

In this section, the numerical models are benchmarked based on experimental results and simulation results using an interface tracking method based on the ALE framework.
All numerical methods are implemented in GeoChemFoam, our reactive transport toolbox, and their implementation is presented in appendices. In order to save on computational time, the equations are implemented using the quasi-static assumption presented in \ref{appenA}, using that, for all our test cases, $\beta Da<<1$ and $\beta Ki<<1$. The solution procedures are presented in \ref{appenB}. The domains are meshed using Adaptive Mesh Refinement (AMR) for the VoS methods and Local Mesh Refinement (LMR) for the ALE methods presented in \ref{appenC}, and adaptive time-stepping strategies presented in \ref{appenD}. Three benchmark test cases are considered. In the first benchmark case, the numerical models are used to simulate the dissolution of a 3D calcite post in a straight microchannel, and the results are compared with experimental results \cite{2017-Soulaine}. Convergence, accuracy and efficiency of the methods are compared. In the second benchmark, the methods are used to simulate dissolution in a 2D micromodel at various dissolution regimes. The accuracy and efficiency of the methods are compared using the ALE method as a reference. The capability of each model to calculate upscaled coefficients is then studied. In the third test case, the iVoS method is used to simulate dissolution in a 3D micro-CT image of Ketton carbonate. The ALE method could not be used in this case due to its computational cost. The accuracy of the methods is compared based on experimental results \cite{2015-Menke} and simulation results \cite{2016-Nunes} using solid balance with a threshold. Two additional simulations at two different dissolution regimes are then run and the accuracy of upscaling laws \cite{2009-Noirel,2017-Menke,2019-Seigneur} are explored.

\subsection{Benchmark 1: calcite post dissolution - comparison with experiment}

In benchmark 1, we dissolve a calcite post with all four numerical methods (ALE, VoS-$\psi$, VoS-$\psi'$ and iVoS) and compare the results with the experimental data from \cite{2017-Soulaine}. In the experiment, an octagonal-shaped calcite post is placed at the center of a straight microchannel and is dissolved by an acidic solution that is flowing past. The experiment and methods are described in detail in \cite{2017-Soulaine}.

Simulations are performed using the VoS methods, and using the ALE method for reference. The domain is a straight microchannel of size 2.67 mm $\times$ 1.5 mm $\times$ 0.2 mm with a calcite post of height 0.2 mm in its center. Images of the post (in \textit{stl} and \textit{h5} formats) are given in the supplementary material. In each case, a cartesian mesh of resolution $\Delta x$=20 $\mu$m is generated, which is snapped onto the calcite solid surfaces for the ALE method (\ref{appenC}). Increased resolutions of $\Delta x=10$ $\mu$m and $\Delta x=5$ $\mu$m at the interface are obtained using LMR or AMR (\ref{appenC}). The initial meshes have respectively 95,900, 109,476 and 160,390 cells for the ALE simulations and 105,000, 114,430 and 174,280 for both of the micro-continuum simulations.

At t=0, a solution of hydrochloride acid is injected from the left boundary at constant flow rate, extrapolated from a zero-gradient pressure velocity field \cite{2016-OpenFOAM}, and constant concentration $c_{i}=0.0126$ kmol/m$^3$. The acid reacts with the calcite surface to produce $Ca^{2+}$ and $HCO_3^-$. However, due to the very low pH, $HCO_3^-$ reacts instantaneously with $H^+$ to give $H_2CO_3^*$. The two reactions can be added and modelled as the single reaction described below.

\begin{table}[!t]
\renewcommand\arraystretch{1.2}
\centering
\begin{tabular}{c|c|c|c}
 Parameter & Symbol & Value & Unit \\
 \hline
Kinematic viscosity & $\nu$ & 2.61$\times10^{-6}$& m$^2$/s \\
Diffusion coefficient & $D$ & $5\times10^{-9}$ & m$^2$/s \\
Inlet flow rate & $Q_i$ & 3.5$\times10^{-10}$ & m$^3$/s \\
Inlet acid concentration & $c_i$ & 0.0126 & kmol/m$^3$ \\
Reaction constant & $k_c$ & 8.9125$\times10^{-4}$ & m/s \\
Stoichiometric coefficient & $\zeta$  & 2 & (-) \\
Calcite molecular weight & $M_{ws}$ & 100 & kg/kmol \\
Calcite density & $\rho_s$ & 2710 & kg/m$^3$ \\
Kozeny-Carman constant & $k_0$ & $10^{-12}$ & m$^{2}$ \\
 \end{tabular}
 \caption{Simulation parameters for Benchmark 1.\label{Table:param1}}
\end{table}

\begin{equation}
 \begin{aligned}
 CaCO_3 + H^+ &\leftrightharpoons Ca^{2+}+ HCO_3^- \\
 HCO_3^-+H^+ &\leftrightharpoons H_2CO_3 \\[0.1cm]
 \hline
 \\[-0.4cm]
CaCO_3 + 2H^+ &\leftrightharpoons Ca^{2+}+ H_2CO_3^* \\
 \end{aligned}
\end{equation}
The simulation parameters are summarized in Table \ref{Table:param1}. Each simulation is run until $t=12000$ s or until all of the solid has been dissolved, whichever happens first. 

\begin{figure}[!b]
\begin{center}
\includegraphics[width=0.85\textwidth]{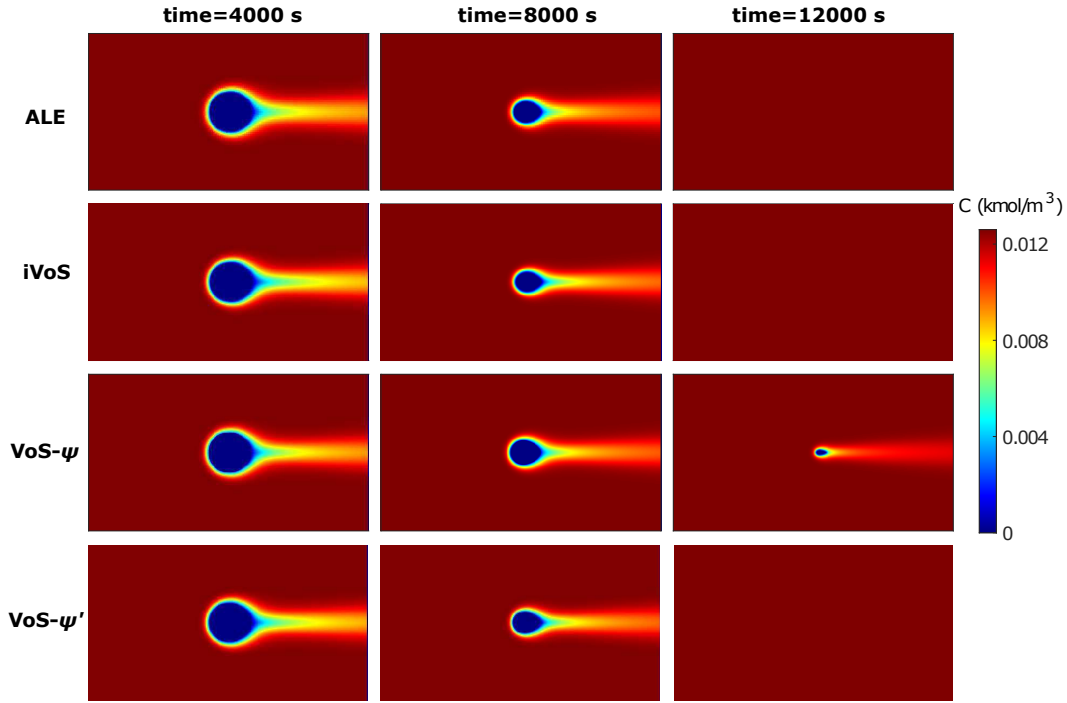}
\caption{Concentration map at various times during dissolution of a calcite post by acid injection using the ALE, VoS-$\psi$, VoS-$\psi$' and iVoS methods.\label{fig:calcitePost}}
\end{center}
\end{figure}

Fig. \ref{fig:calcitePost} shows the concentration map at different times for the different methods, with $c=0$ in the solid phase. We see that for the ALE and iVoS methods, the concentration maps are very similar and the calcite post has fully disappeared at t=12000 s. However, for the VoS-$\psi$ method, the dissolution is delayed and there is still a small but significant volume of calcite at t=12000 s. The VoS-$\psi$' method corrects most of this error and the concentration maps are similar to the one obtained with the ALE and iVoS methods.

\begin{figure}[!t]
\begin{center}
\includegraphics[width=0.99\textwidth]{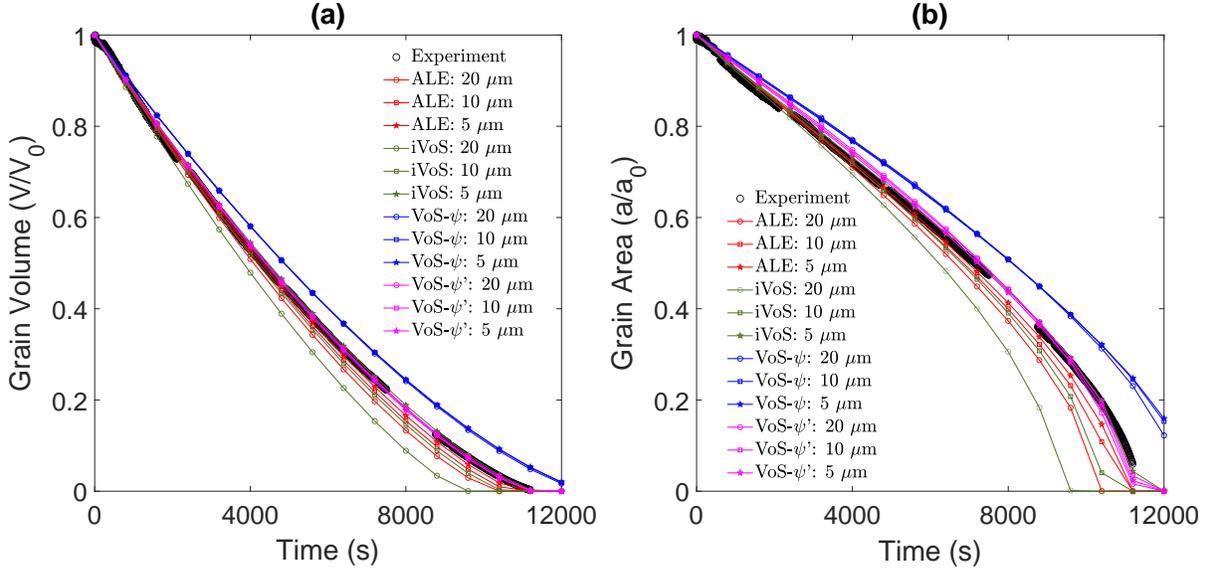}
\caption{Evolution of (a) the grain volume and (b) the grain surface area obtained experimentally (black) and with simulation using the ALE (red), VoS-$\psi$ (blue), , VoS-$\psi$' (purple) and iVoS (green) methods at various mesh resolutions during dissolution of a calcite post by acid injection.\label{fig:calcitePostMass}}
\end{center}
\end{figure}

The results of numerical simulations at different mesh resolutions are compared with experimental results from \cite{2017-Soulaine} in Fig. \ref{fig:calcitePostMass}. We observe that the ALE method converges toward a solution close to the experimental results. The difference between the experimental results and the numerical simulation at mesh resolution $\Delta x=5$ $\mu$m in grain volume and grain area are less than 1\% of the initial values. Although the iVoS method gives a significantly lower grain volume and surface area than the experiment at a resolution $\Delta x=20$ $\mu$m, the results are very similar to the ALE results for $\Delta x=10$ $\mu$m and $\Delta x=5$ $\mu$m. In addition, with the iVoS method as with the ALE method, the difference between the experimental results and the numerical simulation at mesh resolution $\Delta x=5$ $\mu$m in grain volume and grain area are less than 1\% of the initial values.

Although the VoS-$\psi$ method matches the trend of the experiment, it overestimates the grain volume and surface area for all resolutions. The results do not improve as the mesh resolution increases and converge toward a solution with an error of 5\% in the grain volume and 10\% in the grain area. This error is corrected by the surface area correction provided by the VoS-$\psi$' method. At all resolution, the VoS-$\psi$' method gives errors in grain volume and grain area that are less than 1\% of the initial values. During the simulation, $\lambda\left(\varepsilon\right)$ is a value between 8 and 16 that changes as $\varepsilon$ changes. This shows that the VoS-$\psi$ underestimates the overall surface area by a factor between 2 and 4. 

\begin{table}[!t]
\renewcommand\arraystretch{1.2}
\centering
\begin{tabular}{c|c|c|c}
 & 20$\mu$m & 10$\mu$m & 5 $\mu$m\\
 \hline
ALE & 39 & 78 &  147\\
VoS-$\psi$ & 11 & 22 & 62\\
VoS-$\psi'$ & 15 & 30 & 64\\
iVoS & 16 & 31  & 65 \\
 \end{tabular}
 \caption{CPU time [min] obtained with the ALE, VoS-$\psi$, VoS-$\psi$' and iVoS methods at various mesh resolutions during dissolution of a calcite post by acid injection.\label{Table:cpu}}
\end{table}

Table \ref{Table:cpu} shows the CPU times for all simulations. The simulations are significantly faster using the VoS methods, with the iVoS being  approximately 2.5x faster than the ALE for each simulation. The VoS-$\psi$ method is slightly faster than the iVoS and VoS-$\psi$' methods, but this is mostly due to the larger time steps resulting from the lower dissolution rates.

We conclude that all VoS methods are significantly faster than the ALE method. The iVoS provides a result that converges toward a solution with an error less than 1\% when the mesh resolution reaches 5 $\mu$m. The VoS-$\psi$ results in a small error which is due to the reduction of the overall surface area and is corrected in the VoS-$\psi$' method by fitting the constant $\lambda$ for each time-step. The VoS-$\psi$' method gives accurate results for all mesh resolution.



\subsection{Benchmark 2: dissolution regimes in a 2D model}

\begin{figure}[!b]
\begin{center}
\includegraphics[width=0.65\textwidth]{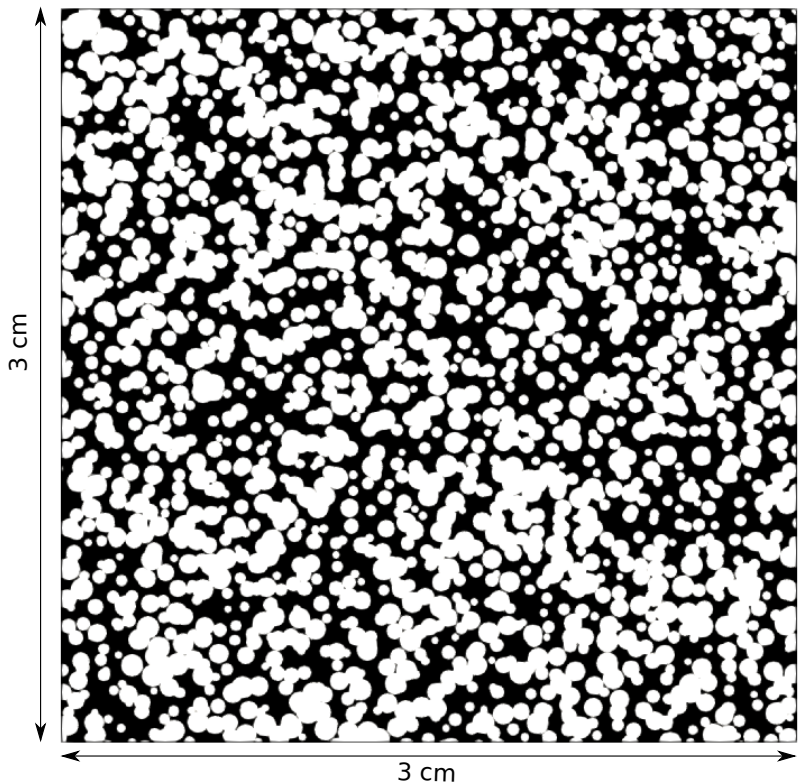}
\caption{Micromodel geometry for Benchmark 2. \label{fig:HM12_6_12}}
\end{center}
\end{figure}

In benchmark 2, we use the three VoS methods to model the various dissolution regimes (i.e compact, wormholes, and uniform) that occur during mineral dissolution in a 2D porous media model. ALE simulations are also run for reference. The model is constructed from a homogeneous domain with discs radius 270 $\mu$m by adding a random deviation of magnitude 270 $\mu$m in disc radius and center position. The micromodel generation code is available open source (\href{https://github.com/hannahmenke/drawmicromodels}{https://github.com/hannahmenke/DrawMicromodels}) and the method is described in \cite{2021-Patsoukis}.  The geometry is presented in Fig. \ref{fig:HM12_6_12} and a high resolution image can be found in the supplementary material.

The domain is meshed with a cartesian mesh with uniform resolution $\Delta x=3$ $\mu$m, which is snapped on the solid surfaces for the ALE method. A band of two cells width is added on each side of the model to avoid dissolution at the boundaries. The final meshes include 1,008,016 cells for the VoS methods and 457,455 for the ALE method.  
The porosity $\phi$ and the permeability $K$ of the full domain can be numerically calculated as $\phi=0.45$ and $K=5.6\times10^{-10}$ m$^2$. 

\begin{table}[!t]
\renewcommand\arraystretch{1.2}
\centering
\begin{tabular}{c|c|c|c}
 Parameter & Symbol & Value & Unit \\
 \hline
Kinematic viscosity & $\nu$ & $10^{-6}$& m$^2$/s \\
Diffusion coefficient & $D$ & $1\times10^{-9}$ & m$^2$/s \\
Inlet acid concentration & $c_i$ & 0.01 & kmol/m$^3$ \\
Stoichiometric coefficient & $\zeta$  & 1 & (-) \\
Calcite molecular weight & $M_{ws}$ & 100 & kg/kmol \\
Calcite density & $\rho_s$ & 2710 & kg/m$^3$ \\
Kozeny-Carman constant & $k_0$ & 1.79$\times10^{-11}$ & m$^{2}$ \\
 \end{tabular}
 \caption{Simulation parameters for Benchmark 2.\label{Table:param2}}
\end{table}

At t=0, acid is injected from the left boundary at constant flow rate, extrapolated from a zero-gradient pressure velocity field \cite{2016-OpenFOAM}, and flows out of the domain from the right boundary at constant pressure. The top and bottom boundaries are no-flow, no-slip conditions. The fluid and solid properties are summarized in Table \ref{Table:param2}. The Kozeny-Carman constant is fitted to obtain the same permeability as in the direct method at $t=0$. For each simulation, the inlet flow rate $Q_i$ and the chemical reaction constant $k_c$ are adapted to obtain the correct $Pe$ and $Ki$, using the pore-scale length $L=\sqrt{12K/\phi}$ as the reference length and the average pore velocity $U=U_D/\phi$ as the reference velocity. The factor 12 is added so that the pore-scale length corresponds to the channel size for an homogeneous bundle of straight channels \cite{2018-Pavuluri}. Four cases are considered that characterize four different dissolution regimes: $Pe=0.01$ and $Ki=0.1$ (compact dissolution), $Pe=0.3$ and $Ki=10$ (conical wormhole), $Pe=1$ and $Ki=1$ (dominant wormhole), and $Pe=10$ and $Ki=0.01$ (uniform dissolution). The simulations are performed until 20\% of the solid has been dissolved.

\begin{figure}[!t]
\begin{center}
\includegraphics[width=0.85\textwidth]{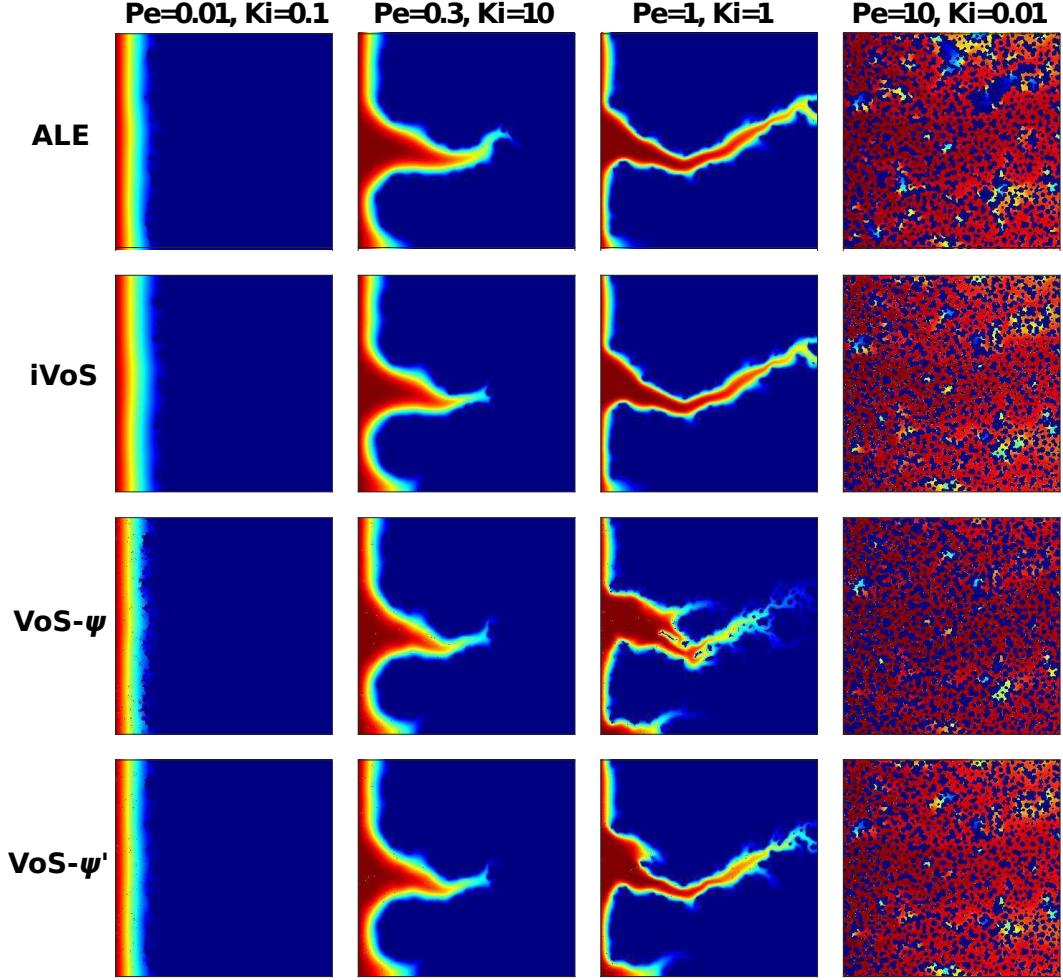}
\caption{Acid concentration map after 20\% dissolution in a micromodel for four different regimes (Benchmark 2) obtained by numerical simulations.\label{fig:micromodelH}}
\end{center}
\end{figure}

\begin{figure}[!t]
\begin{center}
\includegraphics[width=0.9\textwidth]{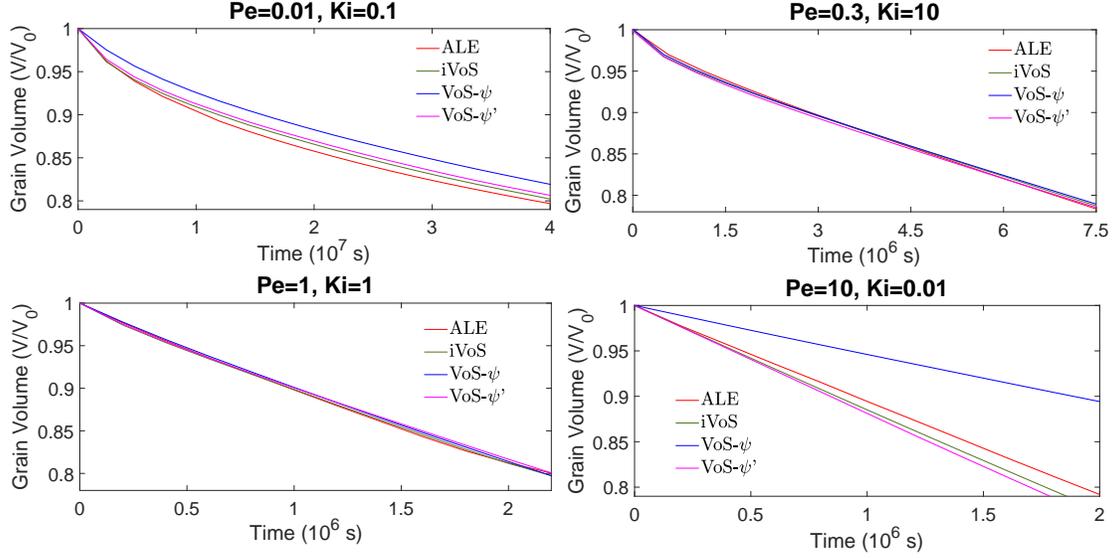}
\caption{Evolution of the solid volume during numerical simulation of dissolution in a 2D micromodel at four different regimes. \label{fig:micromodelV}}
\end{center}
\end{figure}

\begin{figure}[!b]
\begin{center}
\includegraphics[width=0.9\textwidth]{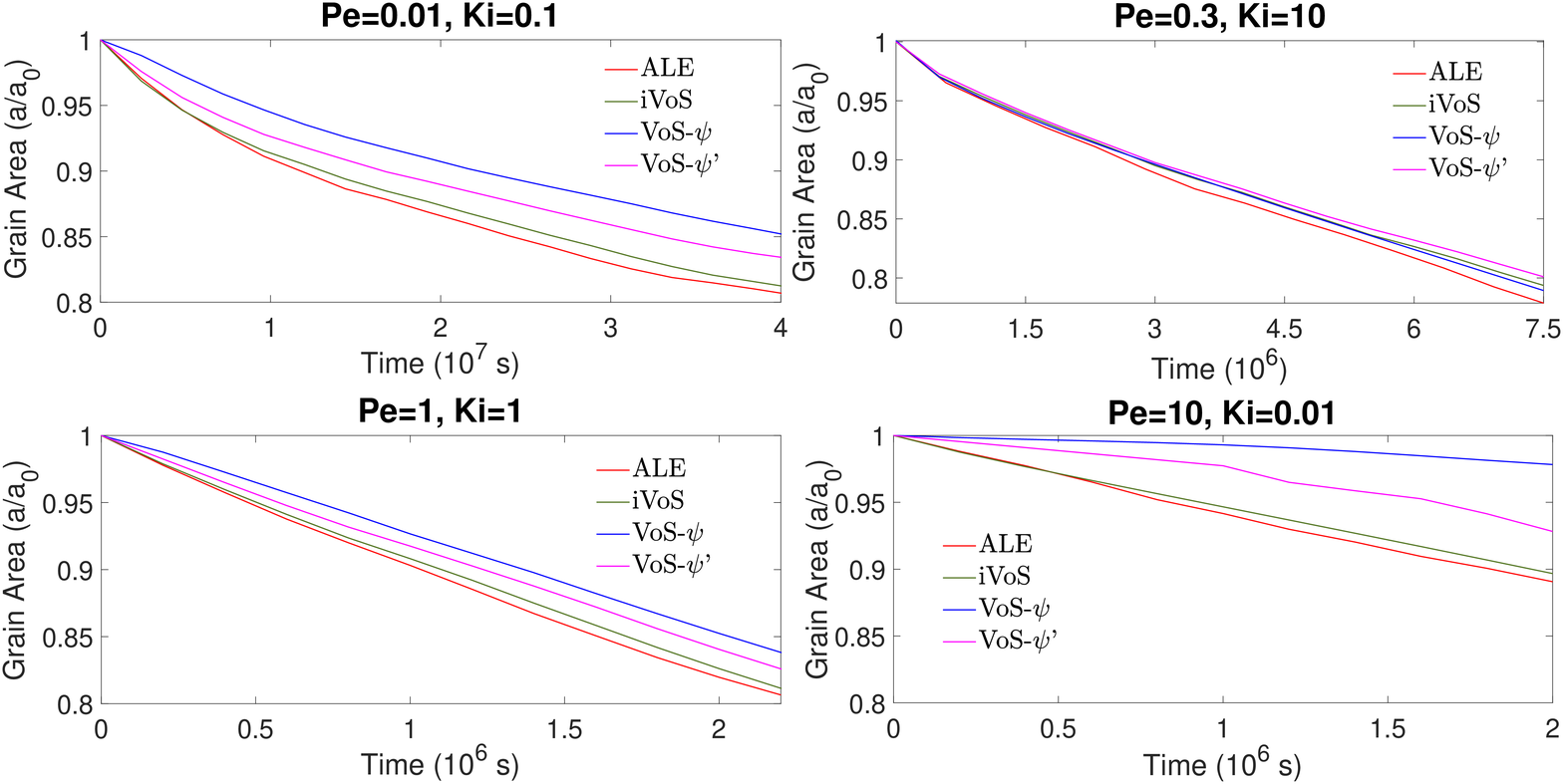}
\caption{Evolution of the solid surface area during numerical simulation of dissolution in a 2D micromodel at four different regimes.\label{fig:micromodelA}}
\end{center}
\end{figure}

Fig. \ref{fig:micromodelH} shows the concentration map at the end of the simulations. We observe that all methods are able to model all regimes qualitatively. In particular, the iVoS method reproduces the ALE results almost exactly. However, there are several inaccuracies in the VoS-$\psi$ method. Due to the use of a centered scheme for the computation of the gradient of $\varepsilon$, the VoS-$\psi$ method results in several grains with incomplete dissolution in the compact and wormhole regimes. This problem can be resolved by applying a decentered gradient, but this is done at the expense of accuracy, as a decentered gradient generates more numerical diffusion. In addition, the wormhole obtained with VoS-$\psi$ for $Pe=1$ and $Ki=1$ is more diffused and ramified that the ones obtained with ALE and iVoS, due to the reduction of reaction rate induced by $\psi$. For the uniform regime, the patterns are almost identical, but the acid concentration obtained with the VoS-$\psi$ method is higher than with ALE and iVoS, suggesting that the reaction rate is lower for VoS-$\psi$. Most of these errors are corrected in the VoS-$\psi$' method, although there are still grains with incomplete dissolution in the compact and wormhole regime, and the wormhole is still slightly more diffuse and ramified for $Pe=1$ and $Ki=1$.

Fig. \ref{fig:micromodelV} and \ref{fig:micromodelA} show the evolution of the solid volume and solid surface area for all regimes obtained with all numerical methods. We observe that the iVoS method reproduces the ALE results with good accuracy for all regimes. For all cases, the solid volume and surface area evolutions obtained with the ALE and the iVoS methods are similar. 
The VoS-$\psi$ method underpredicts the amount of dissolution occurring in the systems compared to the ALE and the iVoS methods. This is particularly true for the cases with $Ki\leq0.1$, i.e. in the compact and uniform regime. In these cases, the concentration gradient on the reactive surface is small and the reaction rate is mostly dependent on the surface area, which is significantly reduced by the localization function $\psi$. For the dominant wormhole regime, although the solid volume evolution obtained with the VoS-$\psi$ method is similar to the ones obtained with the ALE and iVoS methods, the evolution of the solid area is significantly different. This suggests that, although the total amount of dissolution is correctly predicted, it occurs in a slightly different pattern than with the ALE and iVoS methods. The dissolution front is less sharp, and the reaction occurs in the vicinity of the wormhole rather than at the tip of the wormhole as predicted by the ALE and iVoS methods. This leads to a more ramified dissolution pattern, as observed in Fig. \ref{fig:micromodelH}. The VoS-$\psi$' method corrects some of these errors and the overall amount of dissolution is similar to the ALE and iVoS results for all cases. However, the surface area is slightly higher for all cases. This is because, since the dissolution is high when $\varepsilon\approx0.5$ and low when $\varepsilon\approx0$ and $\varepsilon\approx1$, the interface becomes artificially sharp and leads to a larger interfacial area.

\begin{figure}[!t]
\begin{center}
\includegraphics[width=0.99\textwidth]{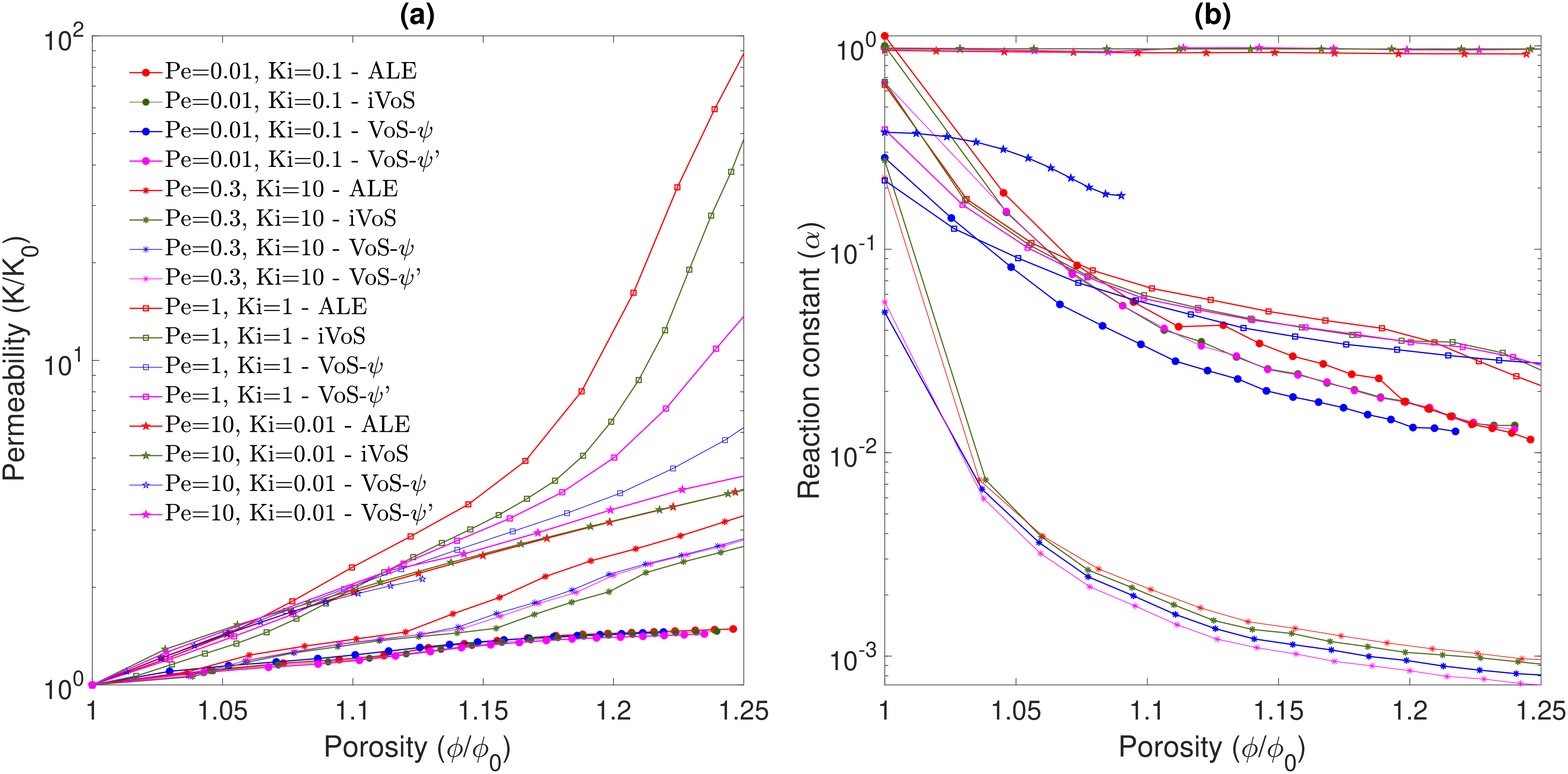}
\caption{Evolution of the permeability and macro-scale reaction constant during numerical simulation of dissolution in a 2D micromodel at four different regimes obtained with the ALE, the VoS-$\psi$ and the iVoS methods.\label{fig:micromodelPerm}}
\end{center}
\end{figure}

Fig. \ref{fig:micromodelPerm} shows the evolution of the permeability $K$ and the correction factor $\alpha$ as a function of porosity for all regimes obtained with the four methods. In all cases, the order of the permeability evolution is similar for the three methods, except for the wormholing regimes, for which the order is 19 for ALE and 17 for iVoS, typical of the dominant wormhole regime, but only 8 for VoS-$\psi$ and 12 for VoS-$\psi$', which is more typical of the ramified wormhole regime. The macro-scale reaction constant is consistently lower with VoS-$\psi$ than with ALE and iVoS, especially for the uniform regime where it is more than twice as small.

\begin{table}[!t]
\renewcommand\arraystretch{1.2}
\centering
\begin{tabular}{c|c|c|c|c}
 & $Pe=0.01$ & $Pe=0.3$ & $Pe=1$ & $Pe=10$ \\
 & $Ki=0.1$ & $Ki=10$ & $Ki=1$ & $Ki=0.01$ \\
 \hline
ALE & 69 & 71 & 46 & 18  \\
iVoS & 5.8 & 6.2  & 4.6 & 4.3 \\
VoS-$\psi$ & 3.8 & 3.8 & 3.3 & 2.7 \\
VoS-$\psi$' & 7.1 & 7.3 & 6.2 & 6.3 \\
 \end{tabular}
 \caption{CPU time (in hours) obtained for all three methods for all four regimes during dissolution in a 2D micromodel.  \label{Table:micromodelCPU}}
\end{table}

Table \ref{Table:micromodelCPU} shows the CPU time for all cases for all three methods. We observe that the VoS methods are between 3 and 12 time faster than the ALE method. The VoS methods are particularly efficient compared to the ALE method for the cases with localized dissolution front, i.e. for the compact and the wormholing regime, for which the ALE method performs a large number of remeshing steps. The VoS-$\psi$ method is slightly faster than the iVoS method, but this is mostly due to larger time steps resulting from the lower dissolution rates. The VoS-$\psi$' method is slightly slower than the iVoS method, which we attribute to a slower convergence of the transport equation due to a more localized dissolution.

We conclude that the iVoS method is capable of modelling all regimes during dissolution in a 2D micromodel and calculating macro-scale coefficients with similar results as the ones obtained with ALE. In addition, the iVoS method is significantly faster than the ALE method. The VoS-$\psi$ method, using $\psi=4\varepsilon_f(1-\varepsilon_f)$ underpredicts the dissolution in all cases due to a reduction of overall surface area, but this error is mostly corrected by using the VoS-$\psi$' method.

\subsection{Benchmark 3: dissolution in a 3D micro-CT image and comparison with experiment}
\label{}

In benchmark 3 we simulate the dissolution of a 3D micro-CT image of Ketton limestone and compare the numerical results to \textit{in-situ} dissolution experiments. The main advantage of the micro-continuum approach is its computational efficiency compared to the ALE method. In this part, we take advantage of this to perform simulations with a VoS method, while the ALE method is too computationally expensive to perform on a 3D image of this size. The iVoS method was selected since it gave the most accurate results in benchmark 2. The experiment is the one conducted in \cite{2015-Menke}, where CO$_2$-saturated brine is injected in a Ketton carbonate core. A core of 4 mm diameter was flooded with a brine solution representative of a typical saline aquifer consisting of 1\% KCl and 5\% NaCl by weight, pre-equilibrated with supercritical CO$_2$ at 10 MPa and 50 °C.  Micro-CT images were acquired after 17, 33, 50 and 67 min.

\begin{figure}[!b]
\begin{center}
\includegraphics[width=0.99\textwidth]{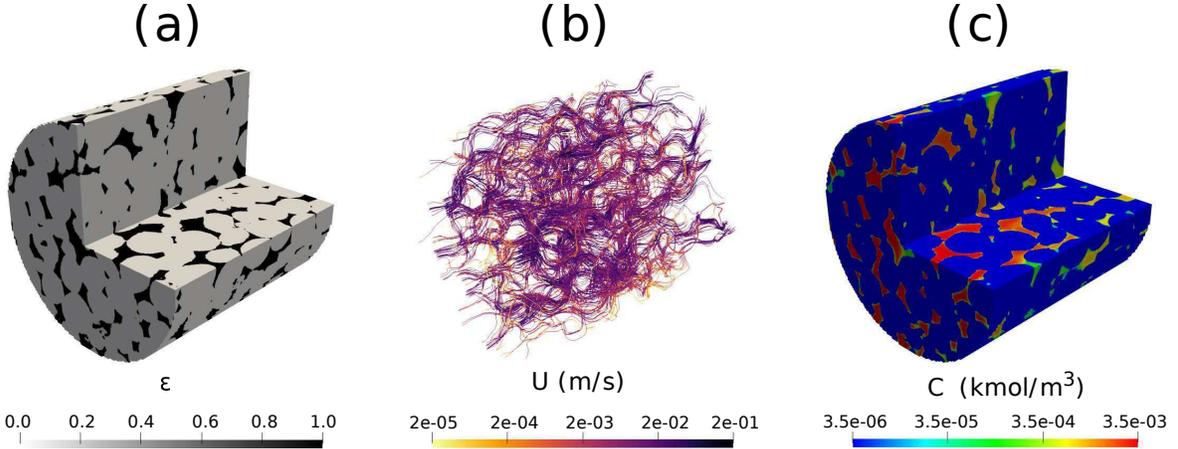}
\caption{Initial condition for simulation of dissolution in a 3D micro-CT image of carbonate: (a) initial local porosity field;(b) Initial calculated velocity field; (c) Initial calculated concentration field ($Pe=190$, $Da=4.4\times10^{-2}$).\label{fig:KettonInit}}
\end{center}
\end{figure}

The simulated conditions are similar to the ones presented in \cite{2016-Nunes}. The sample is a cylinder of radius 1.7mm and length 3.5mm. The image has $911\times902\times922$ voxels with resolution 3.8 $\mu$m, however, we ran it on two-level adpative mesh with maximum resolution of 7.6 $\mu$m to decrease computation time. The initial grid includes 17 million cells. Fig. \ref{fig:KettonInit}a shows the initial porosity field. The velocity field is then initialised by solving Equ. (\ref{Eq:contDBS}) and (\ref{Eq:momentumDBS}) in the domain using $k_0=2\times10^{-13}$ m$^{2}$ which was fitted to obtain a permeability of 16D. The streamlines and corresponding magnitude of velocity are shown on Fig. \ref{fig:KettonInit}b. The reaction rate is described in terms of the concentration of calcium cations Ca$^{2+}$, following $R=k_{eff}\left(C_{eq}-C\right)$, where $C_{eq}$ and $k_{eff}$ depend on the concentrations of $CO_3^{2-}$, $H^+$ and $CO_2$. Because our model only tracks one component, the reaction constant and equilibrium concentration are fitted to match the reaction-limited constant obtained for a flat pure crystal of calcite \cite{2015-Peng} $R=8.1\times10^{-7}$ kmol/m$^2$/s and the initial reaction rate in the experiment $R=8.8\times10^{-8}$ kmol/m$^2$/s. Such rates are obtained using a reactant concentration of 0.0035 kmol/m$^3$ at the inlet and a reaction constant $k_c=2.314\times10^{-4}$ m/s. Fig. \ref{fig:KettonInit}c shows the concentration field in the pores at $T=0$ s obtained by solving Equ. (\ref{Eq:concentrationDBSQS}).

\begin{table}[!t]
\renewcommand\arraystretch{1.2}
\centering
\begin{tabular}{c|c|c|c}
 Parameter & Symbol & Value & Unit \\
 \hline
Kinematic viscosity & $\nu$ & $2.61\times10^{-6}$& m$^2$/s \\
Diffusion coefficient & $D$ & $7.5\times10^{-10}$ & m$^2$/s \\
Inlet flow rate & $Q_i$ & 8.33$\times10^{-9}$ & m$^3$/s \\
Inlet concentration & $c$ & 0.0035 & kmol/m$^3$ \\
Reaction constant & $k_c$ & 2.314$\times10^{-4}$ & m/s \\
Stoichiometric coefficient & $\zeta$  & 1 & (-) \\
Calcite molecular weight & $M_{ws}$ & 100 & kg/kmol \\
Calcite density & $\rho_s$ & 2390 & kg/m$^3$ \\
Kozeny-Carman constant & $k_0$ & 2$\times10^{-13}$ & m$^{2}$ \\
 \end{tabular}
 \caption{Simulation parameters for Benchmark 3.\label{Table:param3}}
\end{table}

The simulation parameters are summarized in Table \ref{Table:param3}. We use for reference velocity the average pore velocity defined as
\begin{equation}
U=\frac{U_D}{\phi}.
\end{equation}
The reference length is associated to the initial permeability so that
\begin{equation}
L=\sqrt{\left(\frac{8K}{\phi}\right)},
\end{equation}
where the constant 8 is added in order to obtain the throat radius for a capillary bundle of uniform size. With these definitions, we obtain $U=5.3\times10^{-3}$ m/s and $L=27$ $\mu$m. Using the diffusion coefficient and reaction constant in Table \ref{Table:param3}, we obtain $Pe=190$ and $Da=4.4\times10^{-2}$.
 
\begin{figure}[!b]
\begin{center}
\includegraphics[width=0.99\textwidth]{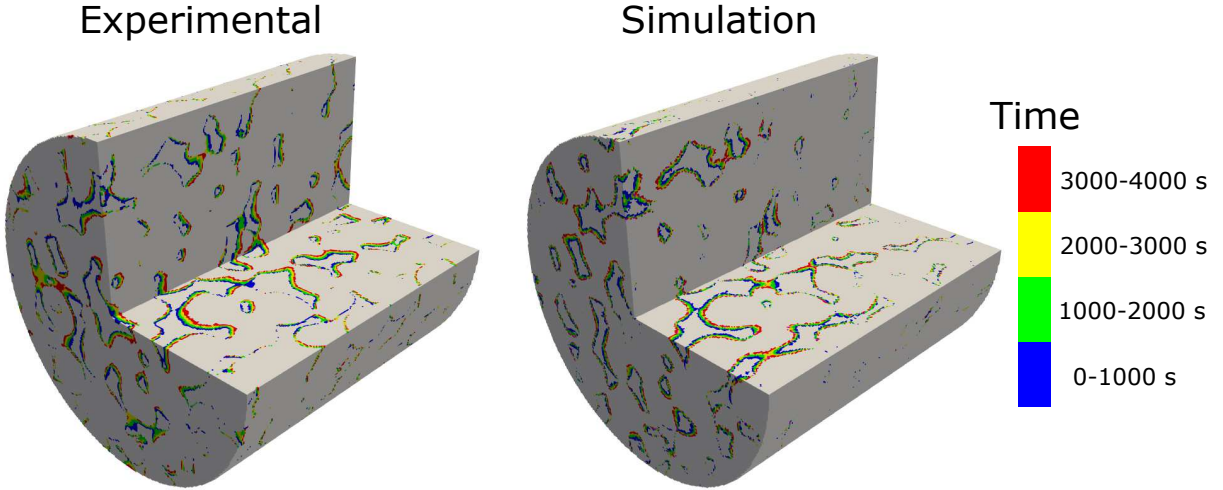}
\caption{Experimental and simulated  results showing the evolution of the dissolution pattern in a 3D micro-CT image of Ketton Carbonate. The colors represent different time intervals.\label{fig:KettonEps}}
\end{center}
\end{figure}

\begin{figure*}[!t]
\begin{minipage}[b]{\textwidth}
\begin{minipage}[b]{0.49\textwidth}
\renewcommand\arraystretch{2}
	\centering
\begin{tabular}{c|c|c||c|c}
 & \multicolumn{2}{c||}{Simulation} & \multicolumn{2}{|c}{Experiment} \\
\hline
 Time & $\phi$ & $K$ (D) & $\phi$ & $K$ (D) \\
\hline
 0 & 0.173 & 16 & 0.173 & 16 \\
1000 & 0.204 & 33 & 0.204 & 37 \\
2000 & 0.228 & 57 & 0.227 & 62 \\
3000 & 0.251 & 89 & 0.248 & 93 \\
4000 & 0.273 & 129 & 0.267 & 129 \\
 \end{tabular}
		\captionof{table}{Experimental and simulated porosity and permeability at different time during dissolution in a 3D micro-CT image of Ketton Carbonate.\label{Table:poroPerm}}
	\end{minipage}
\hfill
\begin{minipage}[b]{0.49\textwidth}
	\centering
\includegraphics[width=0.99\textwidth]{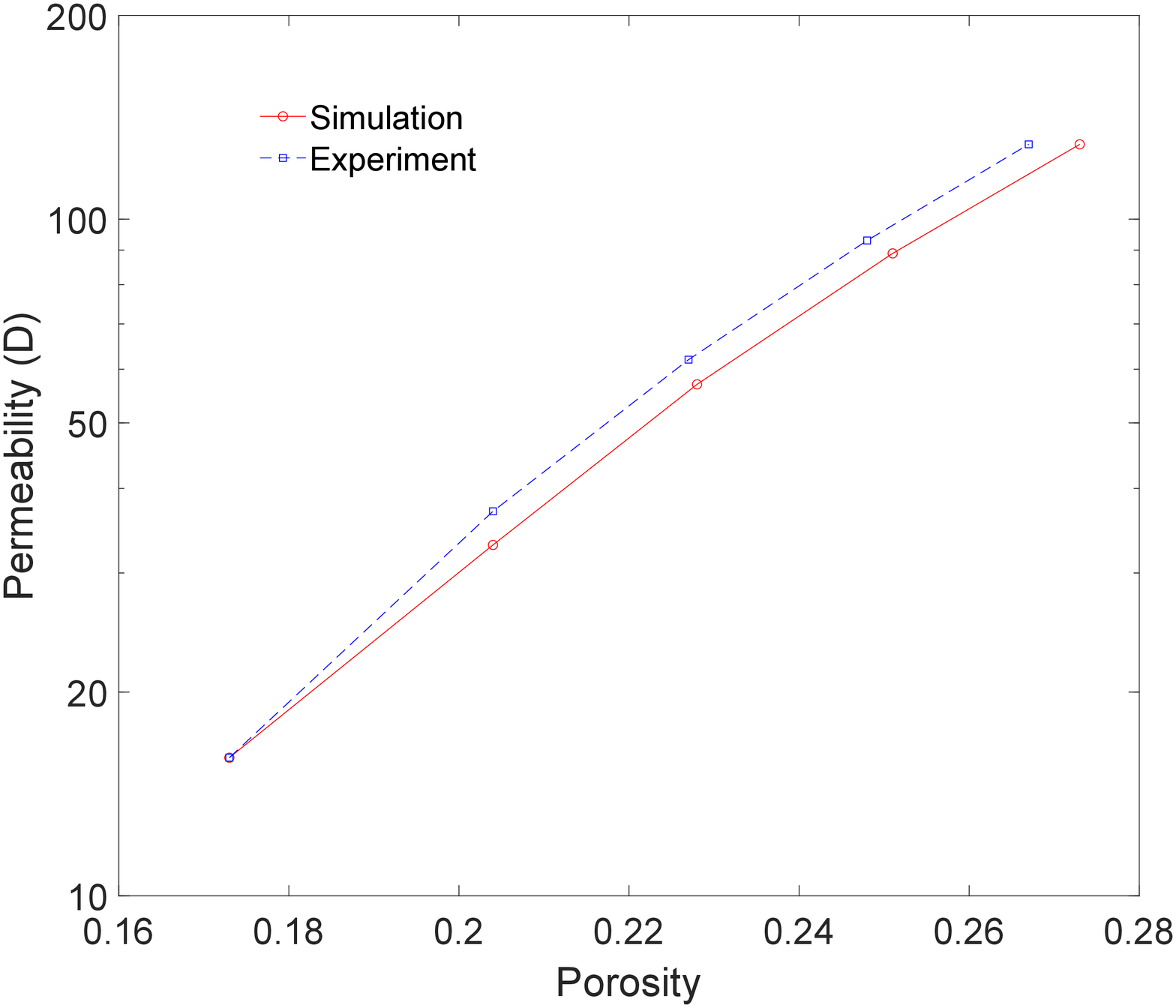}
	\captionof{figure}{Permeability as a function of evolving porosity during dissolution in a 3D micro-CT image of Ketton Carbonate.\label{fig:poroPerm}}
\end{minipage}
\end{minipage}
\end{figure*}

The simulation is run with an adaptive time-step until $t=4000$ s on 128 CPUs using Oracle cloud computing. The total CPU time was 65 hours. Fig. \ref{fig:KettonEps} shows a comparison of the evolution of the dissolution pattern
between experiment and simulation. The colors show the part of the rock that is dissolved for each time interval. We observe that the simulation captures the correct patterns, with pores being enlarged in the direction of the flow. Some expected differences are observed, primarily due to uncertainties in the experimental conditions, segmentation error due to reaction occurring during image acquisition, and the fact that the simulation is performed on a sub-image of the full sample used in the experiment. Furthermore, some of the differences in the local porosity at different times in the experiment are due to imperfect alignment of the images. Table \ref{Table:poroPerm} shows the evolution of the porosity during dissolution for both experiment and simulation. We observe that the porosity is accurately predicted by the simulation until $T>3000$ s, at which time the simulation starts diverging from the experiment slightly. This can be explained by the fact that the simulation is done on a sub-image of the full sample, starting at 2 mm away from the inlet \cite{2015-Menke}, and thus the concentration of acid will have varied as dissolution occured in the unimaged portion of the core.

\begin{table}[!b]
\renewcommand\arraystretch{1.2}
	\centering
\begin{tabular}{c|c|c|c||c|c}
 & \multicolumn{3}{c||}{Simulation} & \multicolumn{2}{|c}{Experiment} \\
 Time (s) & $R_{\Omega}$ (mol/m$^3$/s) & $C$ (mol/m$^3$) & $\alpha$ & $R_{\Omega}$ (mol/m$^3$/s) & $\alpha^*$ \\
\hline
 0-1000  & 0.74 & 1.2 & 0.35  & 0.74 & 0.35  \\
1000-2000  & 0.57 & 1.3 & 0.23 & 0.55 & 0.23 \\
2000-3000   & 0.55 & 1.4 & 0.21  & 0.50 & 0.19\\
3000-4000 & 0.52 & 1.6 & 0.18  & 0.46 & 0.16 \\
 \end{tabular}
		\caption{Evolution of experimental and simulated reaction rate, reactant concentration and macro-scale reaction constant. The experimental macro-scale constant $\alpha^*$ is calculated using the simulated concentration.\label{Table:reactionRate}}
\end{table}

In addition, permeability can be calculated during the simulation. Table \ref{Table:poroPerm} shows the porosity and permeability at different times for the simulation and the experiment and we observe a good correspondence. The permeability is then plotted as a function of the porosity in Fig. \ref{fig:poroPerm}. We see that the permeability in the simulation follows the same trend that of the experiment.
Finally, the average macro-scale reaction rate for each time interval can be calculated as
\begin{equation}
R_{\Omega}=\frac{\rho_s\Delta\phi}{M_w\Delta t},
\end{equation}
for both simulation and experiment. The correction factor $\alpha$ can then be calculated by dividing by the average concentration in the pore space, calculated at the end of the time interval from the simulation results. The values are summarized in Table \ref{Table:reactionRate}. We observe a good correspondence between simulation and experiment. The overall trend of decreasing reaction rates and correction factors observed in the experiment is reproduced in the simulation, albeit with a slightly slower rate. This decrease is not a typical characteristic of the uniform regime, and is an indication that the dissolution might be occuring in the channeling regime, identified in \cite{2017-Menke}.

\begin{figure}[!t]
\begin{center}
\includegraphics[width=0.8\textwidth]{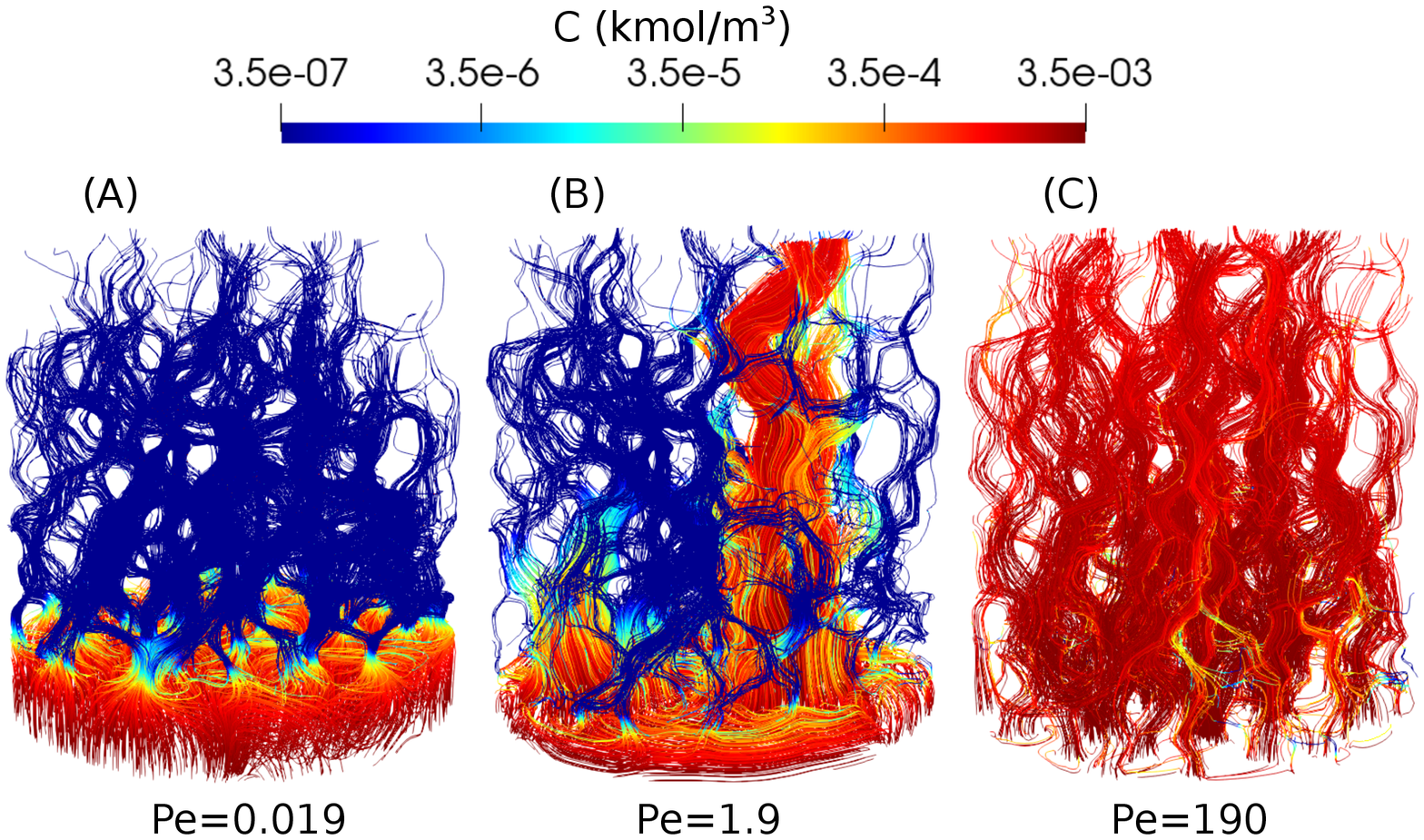}
\caption{Reactant concentration along streamlines during Ketton dissolution in three different regimes, shown when the porosity has reached approximatively 0.28.\label{fig:KettonRegimes}}
\end{center}
\end{figure}

Two additional simulations are run in different regimes by dividing the reaction rate by 100 and 10,000, which gives P\'eclet number of 1.9 and 0.019. The simulations are run on 128 CPUs using Oracle cloud computing until the porosity reaches an approximate value of 0.28. The CPU time was 75 hours for $Pe=1.9$ and 80 hours for $Pe=0.019$.  Figure \ref{fig:KettonRegimes} shows the reactant concentration along reconstructed streamlines at the end of the simulation. At $Pe=0.019$, the dissolution is in the compact regime. The reactant is consumed and dissolves the rock close to the inlet face. At $Pe=1.9$, the dissolution is in the wormholing regime. The reactant penetrates further in the domain and flow instabilities kick in, leading to a preferential dissolution pathway. The streamlines and reactant concentration for $Pe=190$ are also shown in Figure \ref{fig:KettonRegimes} and the dissolution appears to be in the uniform regime.

\begin{figure}[!b]
\begin{center}
\includegraphics[width=0.9\textwidth]{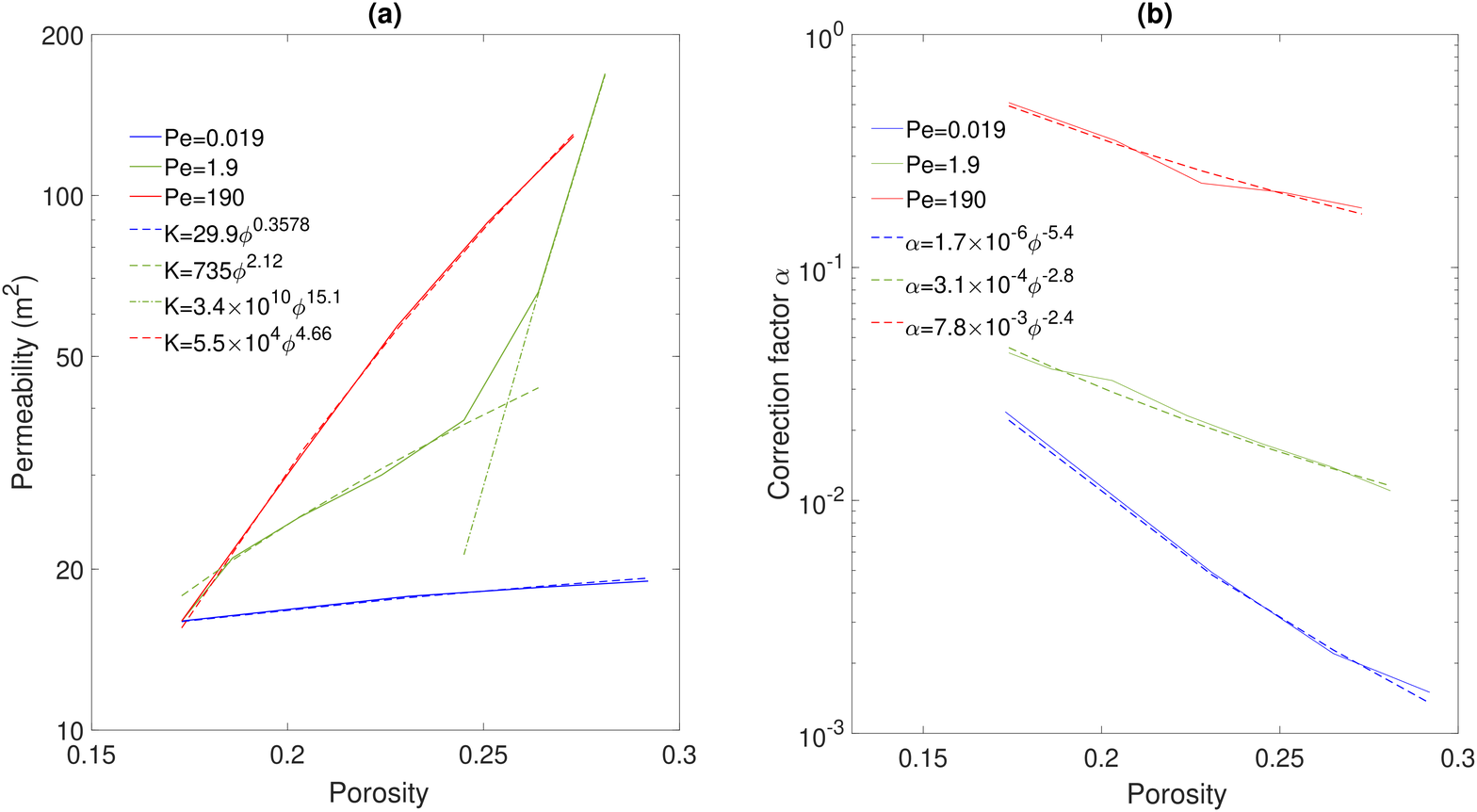}
\caption{Evolution of permeability and correction factor as a function of porosity and corresponding correlations during Ketton dissolution in three different regimes.\label{fig:KettonRegimesCorr}}
\end{center}
\end{figure}

Figure \ref{fig:KettonRegimesCorr}a shows the evolution of permeability as a function of porosity during the simulations for the three regimes. The permeability of the compact and uniform can be fitted with power laws, and we obtain $K=30\phi^{0.36}$ for the compact regime and $K=5.5\times10^{4}\phi^{4.7}$ for the uniform regime. For the wormhole regime, the permeability curve has a strong inflection point around $\phi=0.25$, which corresponds to porosity at reactant breakthrough time. The permeability curve can then be fitted with two power law curves, $K=74\phi^{2.1}$ before breakthrough and $K=3.4\times10^{10}\phi^{15}$ after breakthrough. The orders of the uniform and wormhole regimes correspond those observed in the literature \cite{2017-Menke,2022-Menke}. 

Similarly, Figure \ref{fig:KettonRegimesCorr}b shows the evolution of the correction factor $\alpha$ as a function of porosity during the simulations for the three regimes. For each regime, the evolution of $\alpha$ can be fitted with a power law \cite{2019-Seigneur} and we obtain $\alpha=1.7\times10^{-6}\phi^{-5.4}$ for the compact regime, $\alpha=3.1\times10^{-4}\phi^{-2.8}$ for the wormholing regime and $\alpha=7.8\times10^{-3}\phi^{-2.4}$ for the uniform regime. We observe that the correction factor for the uniform regime is not constant but decreases with a power-law. This indicates that the dissolution might not be in the uniform regime, but could be in the channeling regime identified in \cite{2017-Menke}.

We conclude that the accuracy and computational efficiency of the iVoS method enables simulation of dissolution in a 3D micro-CT image of a real carbonate sample, reproduce experimental results with good precision and can be used to investigate upscaling parameters. 

\section{Conclusion}
We have presented two novel numerical methods, iVoS and VoS-$\psi$', to simulate mineral dissolution in real pore-scale geometries that when compared to existing methods has two main advantages. First, they are based on the micro-continuum approach, and therefore do not require a complex algorithm for interface tracking, nor any special treatment if topological changes occur. Second, the iVoS method calculates a reaction rate based on the divergence of a reactive flux, and thus does not require an interface localization function, which greatly improved its accuracy. The VoS-$\psi$' uses a localization function with a constant that is fitted to ensure that the reactive surface area is conserved globally and therefore avoids most of the numerical errors present with VoS-$\psi$. The advantages of these methods were demonstrated in three benchmark cases. 

In benchmark 1, the iVoS method was used to simulate the dissolution of a 3D calcite post in a straight microchannel, and the results were compared with experimental results and with simulations obtained with the ALE method and with the standard micro-continuum approach based on the VoS-$\psi$ method. We observed that the iVoS method was capable of reproducing the experimental results with similar accuracy but significantly less computational time than the ALE method, while the VoS-$\psi$ method using $\psi=4\varepsilon_f(1-\varepsilon_f)$ showed significantly more error, but this error was corrected by the VoS-$\psi$' approach. All VoS methods were significantly faster than the ALE method.

In benchmark 2, we simulated dissolution in a 2D micromodel and calculated macro-scale coefficients using all numerical methods. Four cases in three different dissolution regimes (i.e. compact, wormholing, and uniform) were considered and we observed a qualitative match between the ALE and iVoS simulations, and a good correspondence between the macro-scale coefficients calculated. The iVoS method was between 4 and 12 times faster than the ALE method. The VoS-$\psi$ method using $\psi=4\varepsilon_f(1-\varepsilon_f)$ was much faster than the ALE method, but underpredicted the dissolution and resulted in inaccurate macro-scale coefficients. Most of these errors were corrected by the VoS-$\psi$' approach, but the results were still slightly less accurate than with iVoS.

In benchmark 3, the computational efficiency of the iVoS method was used to perform a simulation in a 3D micro-CT image of a real carbonate rock (i.e. Ketton) and the simulation results were compared to experimental results. We observe a good correspondence between the experimental and simulated results for the evolution of the total porosity change, total reaction rate and permeability with time. Two additional simulations were performed in different dissolution regimes and the accuracy of correlations for macro-scale coefficients evaluated. We observed that the simulated results could be matched by power-law correlations, and that the coefficients obtained for the permeability in the wormhole and uniform regimes correspond to what has been observed in experiments \cite{2015-Menke,2017-Menke}.
   
In future work, the advantages of our novel approach will be used to perform a large number of simulations in 2D micromodels and 3D micro-CT images that will form a large database for data-driven research. This database will then be used to identify precisely the boundaries between the regimes and decipher the impact of pore-scale heterogeneities. Machine-learning algorithms will also be used to estimate macro-scale coefficients in a similar way that for single-phase flow and reactive transport \cite{2021-Menke,2022-Liu} but extended to predict their evolution with the porosity change. Because the micro-continuum approach can also be used to simulate dissolution at the Darcy-scale \cite{1997-Liu,2000-Ormond,2002-Golfier}, our method can be extended to simulate flow, transport and dissolution in multi-scale porous media \cite{2021b-Patsoukis-Dimou}. Further, the applicabilities of the methods to model precipitation \cite{2021-Yang} will be investigated, Finally, our approach is compatible with the Volume-Of-Fluid and the Continuous Species Transfer methods \cite{2018-Soulaine} for simulation of multiphase flow and multiphase transport with interfacial transfer, so it has the potential to be extended to simulate mineral dissolution during multiphase processes, which is relevant to a number of clean-energy applications, including CO$_2$ storage and geothermal systems \cite{2022-Li}.
  
\appendix

\section{Quasi-static assumption}
\label{appenA}
Dissolution of a solid grain is typically orders of magnitude slower than reactant transport \cite{2012-Szymczak}. Over typical dimensions of pores (10 $\mu$m to 1 mm), the diffusion time scale of the reactant ranges from 1 ms to 10 h. By contrast, to dissolve a layer of calcite 1 mm thick takes at least 1 year \cite{1991-Palmer}. In our dimensionless analysis, this is characterised by the fact that $\beta Da<<1$ and $\beta Ki<<1$. In this case, flow (Equ. (\ref{Equ:momentum})) and transport (Equ. \ref{Eq:concentration})) are assumed to be in a quasi-static state
\begin{equation}
\nabla\cdot\left(\mathbf{u}\otimes\mathbf{u}\right)=-\nabla p +\nu\nabla^2\mathbf{u},\label{Eq:momentumQS}
\end{equation}
\begin{equation}\label{Eq:concentrationQS}
\nabla \cdot \left( c\mathbf{u} \right) = \nabla\cdot\left(D\nabla c\right).
\end{equation}
In this case, the micro-continuum apparoach gives
\begin{eqnarray}\label{Eq:momentumDBSQS}
\nabla\cdot\left(\frac{\mathbf{u}\otimes\mathbf{u}}{\varepsilon_f}\right)=-\nabla \overline{p}_f +\frac{\nu}{\varepsilon_{f}}\nabla^2\overline{\mathbf{u}}-\nu K^{-1}\overline{\mathbf{u}},
\end{eqnarray}
\begin{eqnarray}\label{Eq:concentrationDBSQS}
\nabla\cdot\left(\overline{c}_{f}\overline{\mathbf{u}}\right)-\nabla\cdot\left(\varepsilon D^*\nabla\overline{c}_{f}\right)+\overline{R}_f=0
\end{eqnarray}
and the volume-averaged velocity satisfies the continuity equation
\begin{equation}\label{Eq:contDBSQS}
 \nabla\cdot\overline{\mathbf{u}} = 0.
\end{equation}
The quasi-static assumption allows the models to run with a large time-step controlled only by the velocity of the solid interface (see appendix) to save on computational time.

\section{Implementation}
\label{appenB}
The iVoS and VoS-$\psi$ methods have been implemented in GeoChemFoam \cite{2020b-Maes,2021c-Maes,2021b-Maes,2021a-Maes}. GeoChemFoam is an open-source OpenFOAM\textsuperscript{\textregistered}-based \cite{2016-OpenFOAM} toolbox that contains several additional packages for solving various flow processes including multiphase transport with interface transfer, single-phase flow in multiscale porous media, and reactive transport with mineral dissolution. The full code can be downloaded from \newline \href{www.github.com/geochemfoam}{www.github.com/geochemfoam}.

\begin{figure}[!t]
\begin{center}
\includegraphics[width=0.95\textwidth]{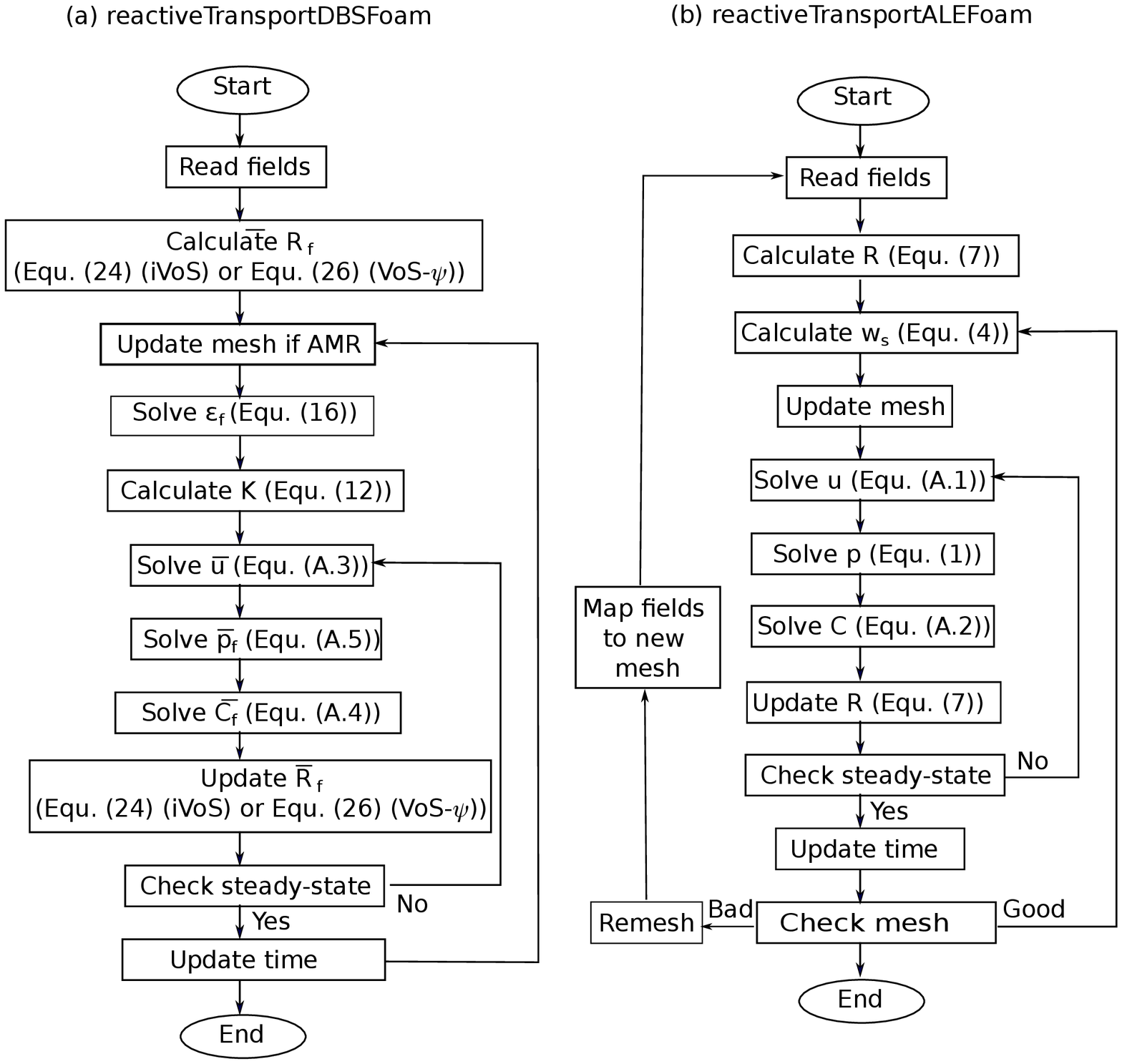}
\caption{Solution procedures for \textit{reactiveTransportDBSFoam} and \textit{reactiveTransportALEFoam} \label{fig:solutionProcedure}}
\end{center}
\end{figure}

The full solution procedure is presented in Fig. \ref{fig:solutionProcedure}a. For each time-step, the new volume of fluid fraction is solved using Equ. (\ref{Eq:mbs}), and the permeability is updated (Equ. (\ref{Eq:perm})). The velocity (Equ. (\ref{Eq:momentumDBSQS})), pressure (Equ. (\ref{Eq:contDBSQS})) and concentration (Equ. (\ref{Eq:concentrationDBSQS})) equations are solved using the Semi-Implicit Method for Pressure Linked Equation (SIMPLE) \cite{1980-Patankar}. The pressure equation is obtained by combining the continuity (Eq. (\ref{Eq:contDBSQS})) and momentum (Eq. (\ref{Eq:momentumDBSQS})) equations. The equations are solved sequentially with under-relaxation factors of 0.3 and iterated until convergence. At the end of each SIMPLE loop iteration, the new volume-averaged surface reaction rate is calculated, using Equ. (\ref{Eq:RfiVoS}) for the iVoS method and Equ. (\ref{Eq:RfVoSpsi}) for the VoS-$\psi$ method. The SIMPLE loop is iterated until steady-state is reached.

In order to compare the iVoS and VoS-$\psi$ methods with an accurate interface tracking method, the ALE method has also been implemented in GeoChemFoam in a separate solver called \textit{reactiveTransportALEFoam} and the full solution procedure is presented in Fig. \ref{fig:solutionProcedure}b. For each time-step, the mesh points are moved with velocity $\mathbf{w}$, which satisfies the Laplace equations with boundary condition (Equ. (\ref{Eq:Ws}))
\begin{eqnarray}\label{Eq:w}
 \nabla\cdot D_m \nabla w_j = 0 \hspace{0.5cm} \text{j=x,y,z}\\
 w_j=\mathbf{w}_s\cdot \mathbf{e}_j \hspace{0.5cm} \text{at $\Gamma$},
\end{eqnarray}
where $D_m$ is the diffusivity of the mesh motion, $w_j$ is the j-directional component and $\mathbf{e}_j$ is the j-directional standard basis vector. These equations will allow the mesh points to track the fluid-solid interface, and the mesh motion is diffused to avoid large volume ratio between neighbor cells. However, the skewness of the mesh can increase and become unacceptably large, which can lead to failure of the transport solver. To avoid this, the mesh quality is checked at the end of each time-step, and if it fails, the domain is fully remeshed and the fields are mapped to the new mesh. To avoid topological errors that can appear when two faces of the same mineral grain overlap, leading to failure of the flow or transport solver, the faces which are fully located in a topological error are eliminated. These collapsing faces are identified by the following condition: a face defined as faceI collapsed if a ray leading from its center following its normal vector pointing toward the solid phase meets another face defined as faceJ at a distance lower than the grid size, and faceI and faceJ do not intersect. Then, a new mesh can be constructed. Following this remeshing algorithm, our numerical simulations are stable and topological errors are eliminated. After the mesh is updated. The velocity, pressure and concentration equations (Equ. (\ref{Eq:cont}), (\ref{Eq:momentumQS}) and (\ref{Eq:concentrationQS})) are then solved using the SIMPLE algorithm with under-relaxation factors of 0.3 and iterated until convergence.

The equations are discretized on a collocated Eulerian grid. The space discretization of the convection terms is performed using the second-order \textit{vanLeer} scheme \cite{1974-vanLeer} while the diffusion term is discretized using the Gauss linear limited corrected scheme, which is second order and conservative. For the iVoS method, the discretization of the reactive flux term (Equ. (\ref{Eq:RfiVoS})) is also done with the \textit{vanLeer} scheme.

\section{Meshing}
\label{appenC}

\begin{figure}[!t]
\begin{center}
\includegraphics[width=0.99\textwidth]{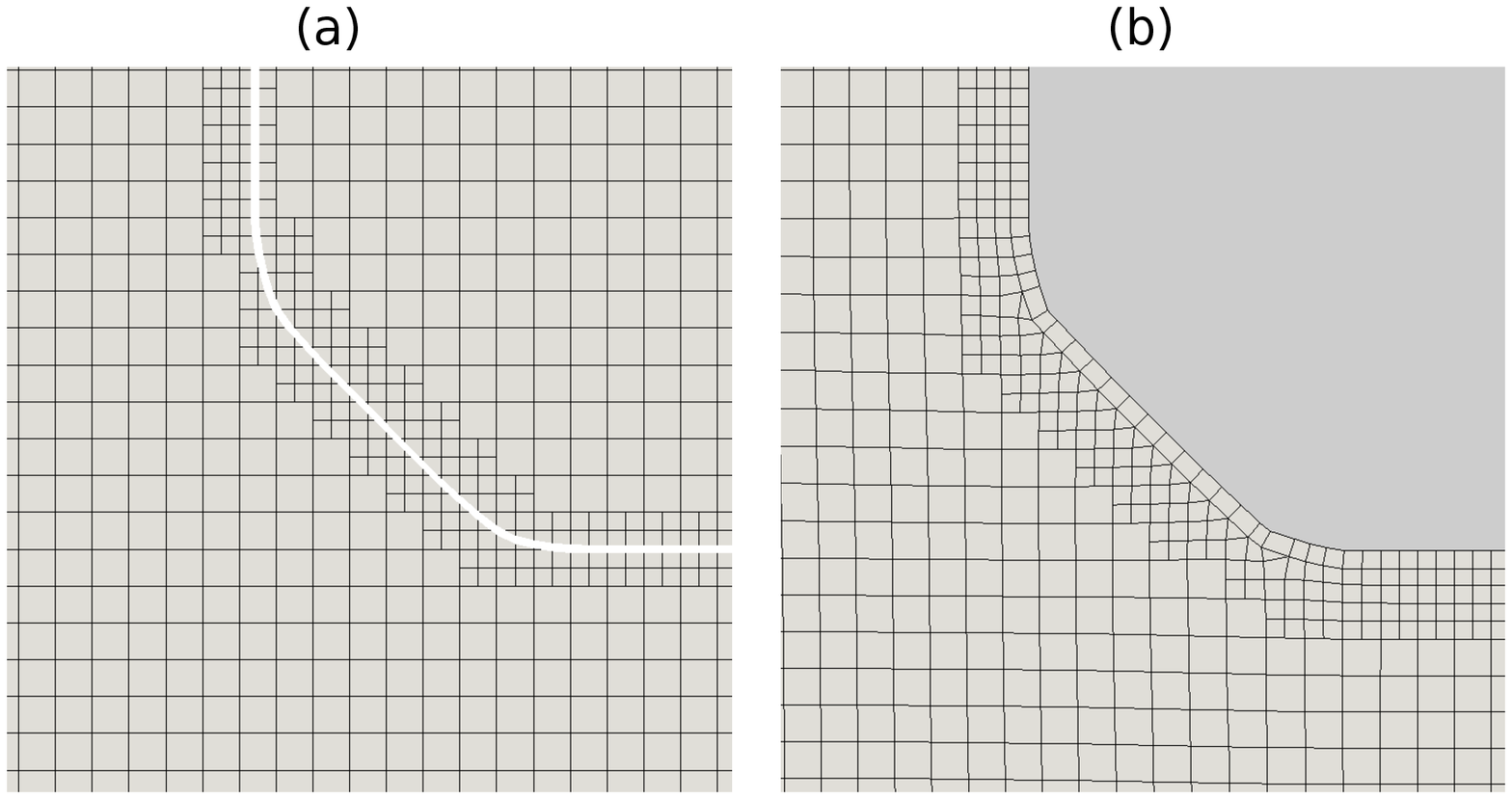}
\caption{Example of adaptive mesh refinement for the micro-continuum approach (a) and local mesh refinement for the ALE method (b) at the surface of an octagonal grain.\label{fig:mesh}}
\end{center}
\end{figure}

For the micro-continuum approach, the solid phase is described using the grayscale value of a \textit{raw} image, encapsulated in \textit{h5} format. A cartesian mesh is generated and the volume fraction of fluid $\varepsilon_f$ in each grid cell is calculated from the image. The volume fraction of fluid in the solid phase is set to a small value $\varepsilon_f=10^{-4}$ to avoid division by zero. An Adaptive Mesh Refinement (AMR) strategy can be used to refine the mesh automatically near the tracked interface (Fig. \ref{fig:mesh}a).
This method splits the computational cell when the volume fraction of fluid satisfies $\varepsilon_{f,min}<\varepsilon_f<\varepsilon_{f,max}$ \cite{2014-Cooke}. The volume fraction of fluid $\varepsilon_f$ is then recalculated for the refined cells with higher resolution before the start of the simulation. During the simulation, we refine cells for which the volume fraction of fluid becomes $\varepsilon_{f,min}<\varepsilon_f<\varepsilon_{f,max}$
and merge them when it becomes $\varepsilon_f \leq \varepsilon_{f,min}$  or $\varepsilon_{f}\geq\varepsilon_{f,max}$. To save on computational time, the mesh is only modified every $n_{ref}$ time-steps. In our simulations, we use $\varepsilon_{f,min}=0.01$, $\varepsilon_{f,max}=0.99$ and $n_{ref}=200$.

For the ALE method, the solid surface is described using an \textit{stl} image. First, a cartesian mesh is generated. Local Mesh Refinement (LMR) can be used to refine the mesh near the solid boundaries (Fig. \ref{fig:mesh}b). The mesh is then snapped onto the solid surface using the \textit{snappyHexMesh} utility \cite{2016-OpenFOAM}, i.e. cell containing solid are then removed and replaced by hexahedral or tetrahedral cells that match the solid boundaries. An additional layer of cells of the same resolution $\Delta x$ is then added around the solid boundary to improve the representation of the solid surface. 

\section{Time-stepping strategy}
\label{appenD}
The simulations are performed using an adaptive time-stepping strategy. For the ALE method, this is done using the mesh Courant-Friedrich-Lewy (CFL) number defined as
\begin{equation}
 mCFL = \frac{\mathbf{w}\Delta t}{\Delta x},
\end{equation}
where $\Delta t$ is the time-step and $\Delta x$ is the mesh resolution. For the micro-continuum approach, the \textit{mCFL} number corresponds a maximum change of $\varepsilon_f$ during a time-step $\Delta\varepsilon_{f,max}$.
Our experience indicates that maximum $mCFL$ number and $\Delta\varepsilon_{f,max}$ of 0.005 offers a good compromise between accuracy, robustness and efficiency, and that is what we are using in this paper.


\bibliographystyle{spphys} 
\bibliography{mybibliography.bib}





\end{document}